\documentclass[sigconf,authorversion,nonacm]{acmart}

\usepackage{soul}

\newtheorem{theorem}{Lemma}

\def\BibTeX{{\rm B\kern-.05em{\sc i\kern-.025em b}\kern-.08emT\kern-.1667em\lower.7ex\hbox{E}\kern-.125emX}}
    
\setcopyright{none}

\begin{document}

%
\title[TDoS ]{Conceptualization and cases of study on cyber operations against the sustainability of the tactical edge}

%

\author{Marco Antonio Sotelo Monge}
\email{msotelo@um.es}
\affiliation{%
    \institution{University of Murcia}
  \institution{Faculty of Computer Science}
  \institution{Campus de Espinardo s/n}
  \city{Murcia, Spain}
  \institution{30100}
  }

\author{Jorge Maestre Vidal}
\email{jmaestre@indra.es}
\affiliation{%
  \institution{Indra}
    \institution{Digital Labs}
  \institution{Avenida de Bruselas, 35}
  \city{Alcobendas, Madrid, Spain}
  \institution{28108}
}

%
\renewcommand{\shortauthors}{Sotelo et al.}

%
\begin{abstract}
The last decade consolidated the cyberspace as fifth domain of operations, which extends its preliminarily intelligence and information exchange purposes towards enabling complex offensive and defensive operations supported/supportively of parallel kinetic domain actuations.  Although there is a plethora of well documented cases on strategic and operational interventions of cyber commands, the cyber tactical military edge is still a challenge, where cyber fires barely integrate to the traditional joint targeting cycle due among others to long planning/development times, asymmetric effects, strict target reachability requirements, or the fast propagation of collateral damage; the latter rapidly deriving on hybrid impacts (political, economic, social, etc.) and evidencing significant socio-technical gaps. In this context, it is expected that tactical clouds disruptively facilitate cyber operations at the edge while exposing the rest of the digital assets of the operation to them. On these grounds, the main purpose of the conducted research is to review and in depth analyze the risks and opportunities of jeopardizing the sustainability of the military tactical clouds at the edge by cyber operations. Along with a 1) comprehensively formulation of the researched problematic, the study 2) formalizes the Tactical Denial of Sustainability (TDoS) concept; 3) introduces the phasing, potential attack surfaces, terrains and impact of TDoS attacks; 4) emphasizes the related human and socio-technical aspects; 5) analyzes the threats/opportunities inherent to their impact on the cloud energy efficiency; 6) reviews their implications at the military cyber thinking for tactical operations; 7) illustrates five extensive CONOPS that facilitate the understanding of the TDoS concept; and given the high novelty of the discussed topics, it 8) paves the way for further research and development actions.
\end{abstract}

%
%




%

%
\maketitle
\pagestyle{plain}

\section{Introduction}
The increasingly digitalization of the military sector is settling the socio-technical foundations towards translating an information advantage, enabled in part by network-centric and fusion warfare paradigms, into a competitive advantage on the joint theatres of operations, secondly supporting the provisioning of intelligence, ISTAR and C2 capabilities thorough the military edge \cite{tdos1}. Characterized as NATO C3 \cite{tdos2} and beyond services, the cyber tactical capabilities may range from logistic, surveillance and reconnaissance, to distributed and federated light algorithmic able to bring automatism as support for decision making to kinetic tactical effectors. But the inventory, provisioning, operation, upkeep and removal of these services is highly technological and data dependant, typically requiring costly data management and distribution procedures \cite{tdos3} hardly affordable on Beyond Line of Sight (BLOS) operational scenarios. In this context, growing concepts like Cloud Computing (CC), Software-Defined Networks (SDN), Network Function Virtualization (NFV) or Self-Organizing Networking (SON) \cite{tdos4} became key enablers for digital tactical capability integration and orchestration at the tactical edge via Tactical Clouds; from which vulnerabilities and cyber attack surfaces are inherited \cite{tdos53} and exposed to cyber adversarial \cite{tdos5}\cite{tdos6}. 

With the motivation of contributing to understand the intersection between the cyber domain and the rest of the operational terrains, the conducted research covered by this paper explores the escalation of a specific family of cyber threats to the Tactical Cloud concept: the Economical Denial of Sustainability (EDoS) attacks; and their evolution on the military context as Tactical Denial of Sustainability (TDoS) actions. The conducted research extends the work preliminary presented to the research community and cyber defence practitioners in \cite{newTDoS104}, compiling the widely received feedback and increasing its original scope from raw technological aspects up to human and environmental aspects. Given the high interest it aroused, the paper additionally includes a section dedicated to the cyber military thinking concerning offensive/defensive denial of sustainability at the edge; and novel illustrative concepts of operation. The following enumerates the main contributions of the conducted research.

\begin{itemize}
\item The state-of-the-art on Tactical Clouds and the impact situations on the cyberspace regarding the rest of kinetic operational domains is widely reviewed and discussed.

\item The term Tactical Denial of Sustainability (TDoS) is formally introduced as evolution of the EDoS concepts to serve tactical objectives.

\item The impact dimensions of TDoS related actuation at technological, tactical, operational and strategic level are studied in-depth.

\item The concepts \textit{Digital Tactical Capabilities}, \textit{Tactical Cloud Provisioning Similarity}, \textit{Maintenance-based Tactical Denial of Sustainability}, and \textit{Deployment-based Tactical Denial of Sustainability} are formulated and discussed.

\item The implications of TDoS against the human sustainability of cyber military tactical operations are analyzed, concluding in the formulation of the paradigms \textit{Enterprise-based Tactical Denial of Sustainability}, \textit{Organizational-based Tactical Denial of Sustainability} and \textit{Individual-based Tactical Denial of Sustainability}.

\item The close relationship between TDoS and the energy efficiency and environmental sustainability of tactical clouds is studied

\item Guidelines for TDoS related offensive and defensive cyber military thinking at the edge are proposed.

\item Five illustrative Concepts of Operations (CONOPs) are detailed (hybrid, proxy and symmetrical war scenarios) as cases of study, which highlight the vertical/horizontal propagation and effect of TDoS at different operational contexts, including digital tactical capability supplying and their potential in exposing the Infrared (IR) signature of stealthy actors   .
\end{itemize}

The paper is organized into nine sections, being the first of them the present introduction. Section II describes the emerging paradigms on tactical cloud computing, as well as the basis of the EDoS attacks. Section III formalizes the TDoS concept, related attack surfaces and impact dimensions. Section IV discusses the potential of TDoS against jeopardizing the human sustainability of tactical clouds. Section V explores the energy efficiency and environmental impacts of TDoS. Section VI addresses the implication of TDoS in terms of cyber military thinking at the tactical Edge. Section VII illustrates five complementary CONOPs where TDoS became a key tactical military enabler. Section VIII widely discusses the CONOPs presented at Section VII. Finally, Section IX summarizes the reached conclusions and foreseen research actions.

Before starting reading the rest of this paper, \ul {the authors want to let the reader know that the vision about military thinking, military operations on the cyberspace and tactical cloud operation does not correspond in any way to the transposition of the doctrine of any particular army nor coalition force, being a vision of their own, resulting from the experience acquired after several years of experience in related sectors. The concepts of operation that are introduced are fictitious scenarios with a mere didactic purpose aiming on facilitating the understanding of the TDoS concept and its potential on military situations. Any deductible Rules Of Engagement (ROE), \textit{jus ad bellum}, or \textit{jus in bello} premise do not intentionally align with any particular international agreement.}

\begin{figure*}[t]
  \centering
  \includegraphics[width=0.6\linewidth]{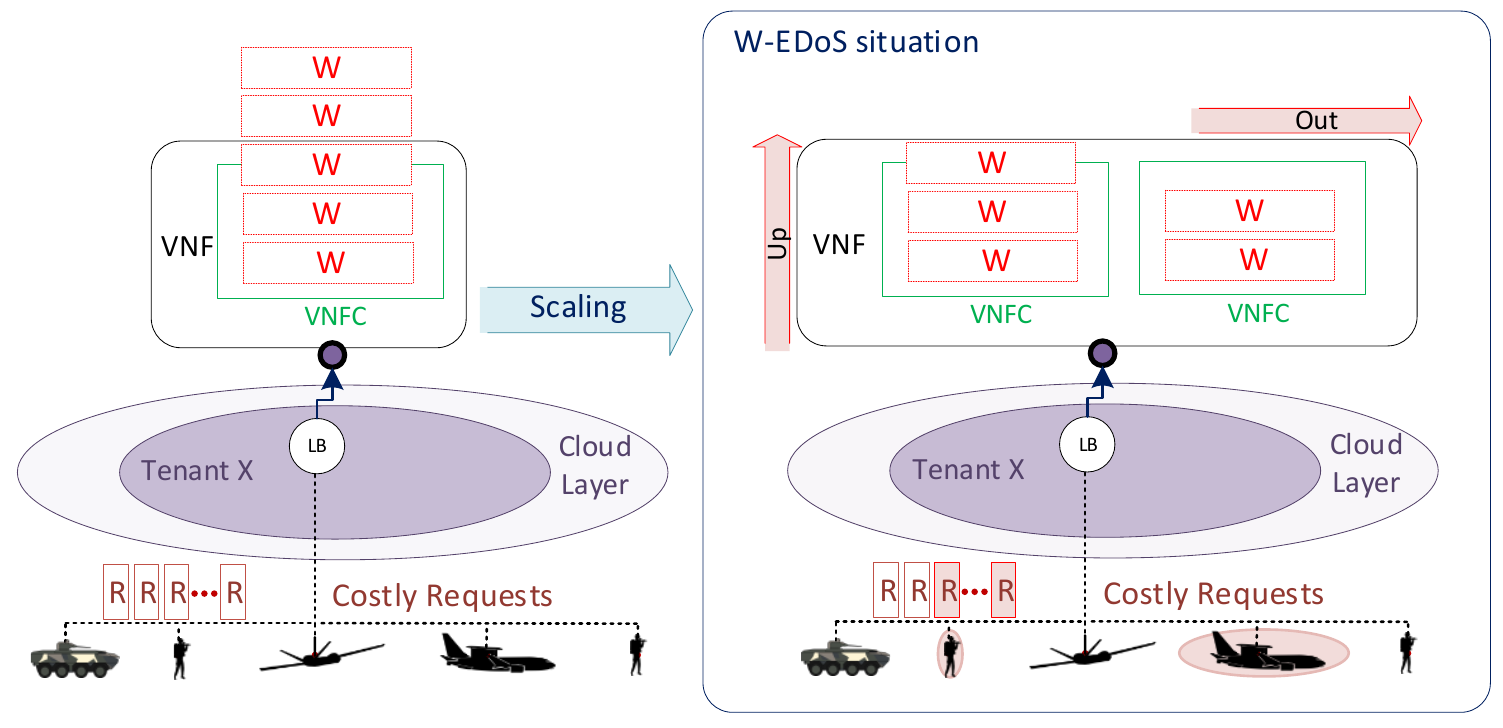}
  \caption{Example of W-EDoS attack on a tactical cloud}
  \label{figW}
\end{figure*}

\begin{figure*}[t]
  \centering
  \includegraphics[width=0.6\linewidth]{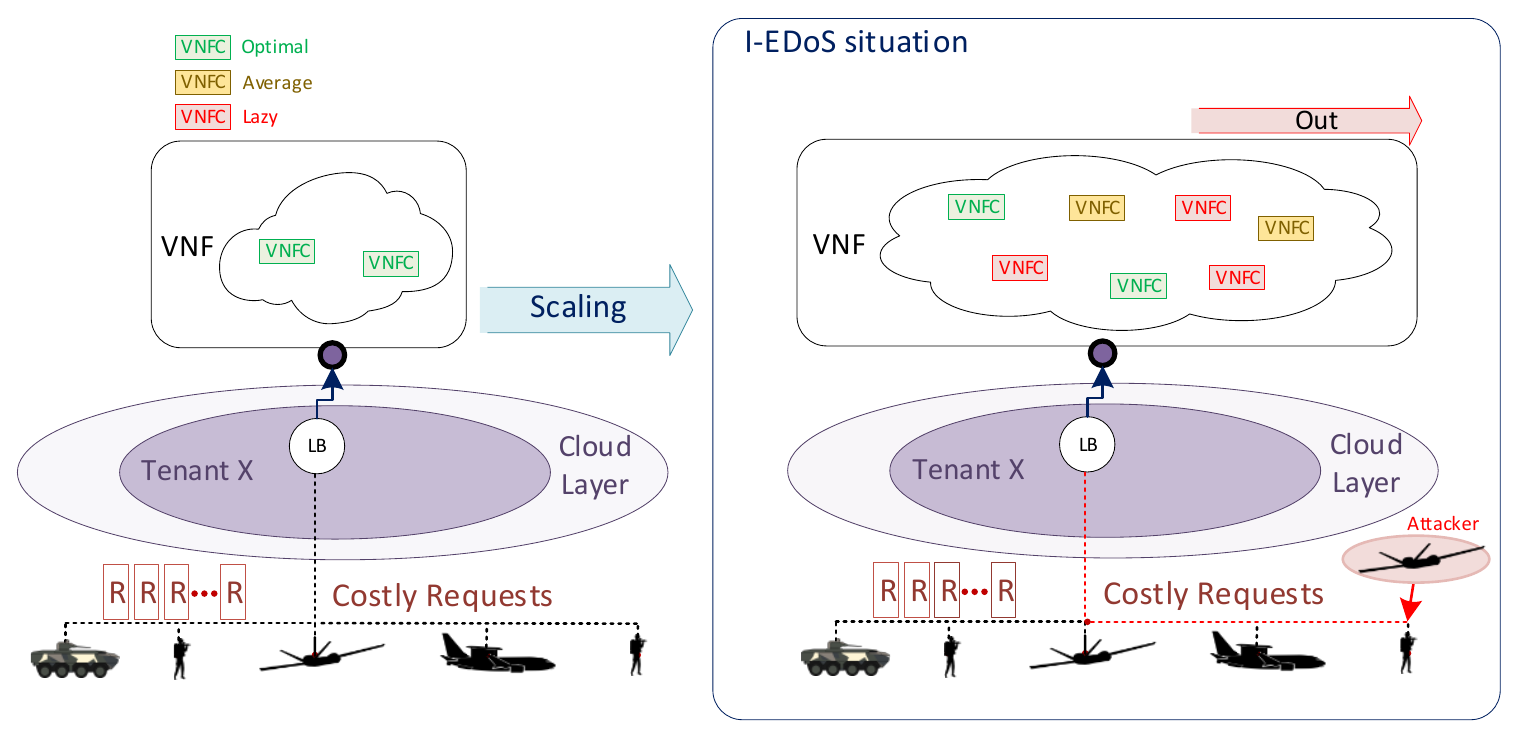}
  \caption{Example of I-EDoS attack on a tactical cloud}
  \label{figI}
\end{figure*}

\section{Background}
This section describes the Tactical Cloud Computing paradigms, including its essential concepts, technological dependencies, and foreseen challenges/gaps. On the other hand, the Economic Denial of Sustainability (EDoS) threats are reviewed, emphasizing their key element and features in the scope of being adapted to BLOS tactical environments.

\subsection{Tactical Clouds}

The Tactical Cloud concepts born in the need to meet the critical communication needs at the tactical edge of military BLOS operations \cite{tdos34}, preliminarily bringing capabilities that among others shall support mobility, heterogeneous waveform, routing policy, or the scalability of the computational resources thorough the edge. Similarly to MANETs/FANETs, military BLOS communications are defined under the condition of no base station, for which edge computing and decision-making were typically combined \cite{tdos36}. Because of their high mobility, their nodes can be deployed quickly, temporarily organize and have strong ability to resist destruction and self-healing, being characterized by enabling distributed peer-to-peer tactical networks, allowing cognitive/dynamical changes in topology and information multi-hop transmission; but also inherited some of the key drawbacks of  BLOS deployments, like energy or bandwidth limitations \cite{tdos35}.
Because of this, they embrace emerging paradigms like Ground Centric Networking (GCC), Cloudlets and Self-Orgaizing Networking (SON). The first provides resilient, bandwidth-efficient communication to users participating in interest groups which have numerous applications to military tactical edge networks due that army units are organized in a hierarchical manner \cite{tdos37}. Cloudlets can be viewed as a “data centers in a box” that “brings the cloud closer” to the battle \cite{tdos38}. The proximity of a cloudlet to its associated mobile devices is the key to its value in hostile environments \cite{tdos39}. Their original motivation was to reduce end-to-end latency of cloud offload from mobile devices for applications that are both resource-intensive and latency sensitive, additionally posing less cyber-physical attack surfaces and sources, enabling self-organization and facilitating the ICT heterogeneity. On the other hand, Self-Organizing Networks (SON) arose with the goal of moving forward from traditional manual management processes towards an automatic and dynamic perspective \cite{tdos52}, for which Software Defined Networking (SDN) reduced the network management complexity by decoupling the control channels, and Network Function Virtualization (NFV) allowed decoupling the software implementation of Network Functions (NF) from the underlying hardware, providing flexibility in the management of the network resources \cite{tdos39}. When deployed at the tactical edge, these capabilities  facilitate the deployment of cloud infrastructure to enable distributed mission command leveraging common software/hardware solutions, and provisioning C3 services thorough conventional and BLOS operational circumstances \cite{tdos40}.
As highlighted in \cite{tdos41}, modern conflict will increasingly rely on advanced information technologies, highlighting four essential pillars: networks, combat cloud, multi-domain battle and fusion warfare. In this context, the Tactical Cloud improves the capabilities for acquiring situational awareness, makes long-range engagements more practical, moves advanced surveillance, data collection and AI to the tactical edge, and supports the development of multi-domain Common Operational Pictures (COPs). However, and as it will be discussed thorough this paper, tactical clouds inherit part of the attack surface of the technological ecosystem that integrates, as is the case of cybersecurity issues and several threats against the cloud sustainability, as is the case of  EDoS.

\subsection{Economic Denial of Sustainability}

The Economical Denial of Sustainability (EDoS) attacks are novel situations preliminarily hypothesized by Hoff \cite{tdos42}\cite{tdos43}, which at the dawn of the cloud computing ecosystem, echoed of potential hostile situations where the attacker may attempt to impact on the cloud computing sustainability by maliciously enforcing economic costs derived from increasing the economic costs derived from both maintenance and provision of the services offered \cite{tdos3}. Despite their novelty, EDoS caught the attention of the research community, which framed their modus operandi as part of the Reduction of Quality (RoQ) \cite{tdos45} and Fraudulent Resource Consumption (FRC) threats \cite{tdos46}\cite{tdos47}\cite{tdos48}. In \cite{tdos7}\cite{tdos8} the problem of EDoS was reviewed in the context of the Self-Organizing Network (SON) paradigm, which presented one of the studies closer to the TDoS paradigms. Accordingly, SON deployments may be jeopardized for causing Workload-based EDoS (W-EDoS) and Instantiation-based (I-EDoS), W-EDoS being caused by maliciously forcing huge workloads (e.g. execution of high complexity algorithms, exploitation of software vulnerabilities, etc.) that require the vertical-escalation of the SON capabilities (see Fig. \ref{figW}); while I-EDoS is caused by the massive deployment of unnecessary (referred to as lazy) VNFs, which was demonstrated by poisoning the telemetric services of the SON Network Function Virtualization Infrastructure (NFVI) (see Fig. \ref{figI}). As highlighted in \cite{tdos46}, beyond the economic impact, EDoS attacks entail several cross-cutting situations, which among others concern the computational capabilities of the cloud, performance, latency, connectivity, availability; and from the socio-technical aspect negatively affect the trust between customers and Digital Service Providers (DSP) (in both direction).
Although in \cite{tdos46} was demonstrated that the implementation of cybersecurity measures based on predicting the behavior of the protected system, constructing adaptive thresholds, and clustering of VNFs instances based on productivity, were effective enough to reveal EDoS threats \cite{tdos54}, their prevention, detection, mitigation and attribution still entail important research challenges, to which is added that the bibliography does not include a large collection of publications focused on the defense against EDoS threats. The studies that address this problem usually assume metrics at network-level, usually confusing features for EDoS identification with those that typically detect flooding-based DDoS behaviors \cite{tdos49}\cite{tdos50}\cite{tdos51}.

\section{Tactical Denial of Sustainability}
Bearing in mind the semantical similarities between Flash Crowds, DoS/DDoS and EDoS threats, in \cite{tdos7} a discriminative criterion was introduced: the CRoWN indicators. Accordingly, the key differentiation between those situations rely on the four parameter directly linked to the client-server model and the distributed provisioning of resources: clients ($C$), requests ($R$), workload that entails their resolution ($W$) and the network functions necessary for their processing ($NF$). Accordingly, W-EDoS are characterized by unexpected workloads ($W$), while I-EDoS are characterized by the suspicious deployment of new functionalities on the distributed environment. As highlighted in Section 2.2., these threats operate on the cyberspace, attempting to achieve technological goals. According to the military theory, they may entail potential threats at the Technological level of war. Beyond the original scope of the EDoS attacks, the TDoS situations target to impact on the Tactical level of war \cite{tdos9} by jeopardizing the sustainability of the capabilities provisioned by the Tactical Clouds, so they embrace tactical-level actuations focused on jeopardizing the decisions and actions that shall originally create advantages when in contact with or in proximity to the enemy. Although this research hypostatizes about the vertical propagation of EDoS to TDoS, other tactical level situations may derive on TDoS.

\begin{figure}[t]
  \centering
  \includegraphics[width=1.0\linewidth]{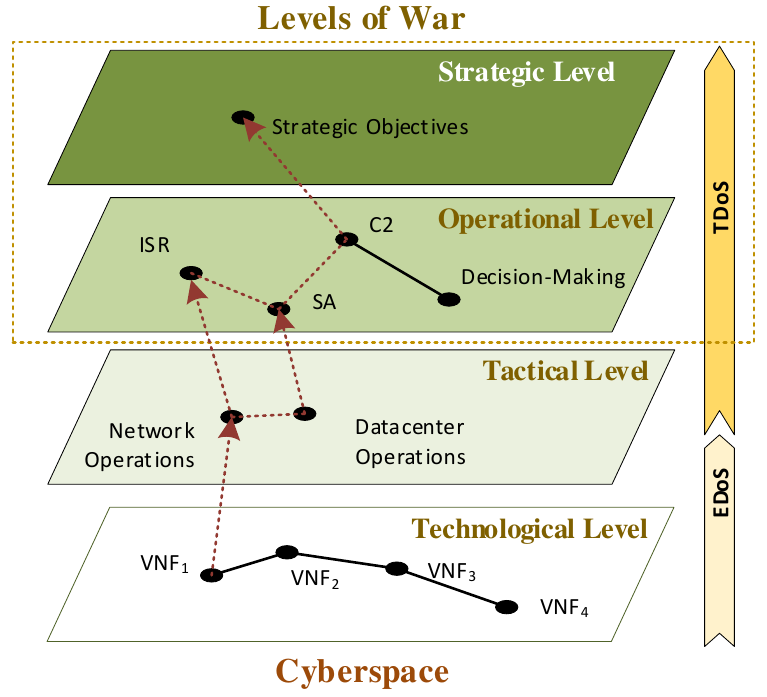}
  \caption{Example of vertical propagation of a TDoS threat}
  \label{fig1}
\end{figure}

An illustrative example of EDoS to TDoS is illustrated in Fig, \ref{fig1}. Accordingly, a TDoS threat aimed on reducing the ISR, C2 and Situational Awareness capabilities supplied by tactical nodes, having the potential effect of vertically impacting on the operational capabilities and strategical objectives that drive a military mission. With this purpose, the attacker propagates a vertical threat from the network and datacenter business processes to the tactical plane. The impact on the tactical level is inferred by jeopardized Virtual Network Functions (VNFs) deployed at the Tactical edge, which implement the tactical services \cite{tdos10}\cite{tdos11}. These services feed operational intelligence, C2 and join planning capabilities, from which the fulfillment of the strategic objectives depends. In this context, by achieving a conventional I-EDoS situation \cite{tdos8}, the attacker forced the instantiation of redundant VNFs, with among others, heavily impact on the energy efficiency of the mobile tactical infrastructure that enables network and datacenter operations. They support the IST, SA, C2 and Decision-making services deployed at the Tactical Edge, which usability will be reduced as the redundant VNFs pointless consume energy.   Consequently, and as depicted in the example, the motivation of EDoS and TDoS is different. The first of them attempts to infer a computational overhead in the VNFs, their supportive infrastructure and their related business processes; and the second attempts to reduce the usability of the tactical services deployed at the edge, which is achieved by depleting the energy supply of the Tactical Cloud enablers. The connection between EDoS and TDoS is the energy depletion, where EDOs became the ``cause'' and TDoS the effect. As commented before, it is expected that non EDoS related attack vectors lead to TDoS situations, so they entail different but related families of threats. So TDoS is a Tactical-level threat potentially unchained by from technological-level situations up to horizontally propagated tactical level threats, being significantly beyond the scope of EDoS.

\subsection{Digital Tactical Capabilities}

The digital tactical capabilities entail the primarily attack surface fo the TDoS threats. The conducted research assumed that these capabilities when deployed on tactical nodes resemble a lifecycle similar to that of the VNFs \cite{tdos12}, which among others shall enable rapid deployment of new functionalities, the scalability of the existing services, shall be interoperable and increase the tactical flexibility (at all C3 Taxonomy levels); from the back-end capabilities to the Operational context. However, a digital tactical capability may be habilitated by various different VNFs, services, or other technological/tactical assets, which lifecycle at least shall embrace some of the following phases:

\begin{itemize}
\item \textbf{Onboarding}. End-to-end provisioning of tactical capabilities from dual-use inventories and/or repositories beyond the operational environment, to the tactical edge. This requires a full technological interoperability, where the onboarded capabilities that support the feature are properly harmonized regarding the preliminarily deployed CIS services and assets.

\item \textbf{Deployment}. The onboarded capabilities require the installation, configuration and coordination of their related CIS enablers. On the other hand, they shall coordinate and cooperate with the rest of the C3 capabilities covered by the Tactical Cloud, all of this being covered at the deployment stage.

\item \textbf{Surveillance}. In order to identify potential risks and malfunctions, the deployed tactical capabilities should be real-time monitored and verified. This is also important to acquire tactical awareness and facilitating the tactical onboarding and deployment of further C3 capabilities.

\item \textbf{Adaptation}. The Tactical Cloud shall enable scalability and flexibility enough to guarantee the proper provisioning and operation of the tactical capabilities. At the adaptation stage, correction, optimizations, and scalability/flexibility issues are continuously conducted, which occur reactive/proactively and driven by automatisms or human intervention \cite{tdos55}. 

\item \textbf{Undeployment}. The deployed tactical services may require to be removed after a certain time period expires, when the accomplished their purport (are they are not needed for the rest of the supported missions phases), or when operational circumstances require it (e.g. as result of enforcing tactical/technological CoAs.
\end{itemize}

\subsection{Conceptualization}

The rapid provisioning of ondemand digital tactical capabilities thorough the tactical edge is one of the key features of a Tactical Cloud, which processes usually are close to cyber expeditionary warfare actuations (in particular, when operating at BLOS). Under the tactical expectations caused by a Rapid Deployment Force \cite{tdos13}, digital tactical capabilities shall be onboarded and deployed at the distributed tactical nodes. In these grounds, the holistic Tactical Denial of Sustainability  concept is formally defined as highlighted in Lemma 2, which depends of the Tactical Communications Similarity relationship formalized in Lemma 1.

\begin{theorem} \textit{Tactical Cloud Provisioning Similarity}\\
\label{lemma-1}

In analogy with EDoS situations, it is expected that during a TDoS attack the digital tactical capabilities pose great resemblance regarding the normal and legitimate circumstances \cite{tdos14} If this is not the case, the tactical disruptions may be caused by alternative operational conditions (denial of service attempts, massive demand by legitimate circumstances, flash crowds, etc.). So let the state $A$ that depicts the normal operability of the tactical communication networks, the State $B$ only can resemble a TDoS situation if: $nA_A \sim nA_B$, $nA_A(t) \sim nA_b(t)$, $nT_A \sim nT_b$ and $nT_A(t) \sim nT_b(t)$, where:
\end{theorem}

\begin{itemize}
\item $nA$: total number of tactical actions that require certain digital tactical capability.
\item $nA(t)$: distribution of the number of tactical actions that require certain digital tactical capability as a mission progress.
\item $nT$: total number of tactical capabilities that requests certain digital tactical capability (horizontal dependencies).
\item $nT(t)$: distribution of the number of tactical capabilities that requests certain digital tactical capability as a mission progress (horizontal dependencies). 
\end{itemize}

Hereinafter this relationship is referred to as \textit{Tactical Cloud Provisioning Similarity} (abbreviated as TPS), so if all the aforementioned conditions are satisfied, it is possible to state that A and B are tactical-based similar (TPS(A,B)).

\begin{theorem} \textit{Tactical Denial of Sustainability}\\
\label{lemma-2}
Let the state A that depicts the normal and expected behavior of a capability delivered by services enabled by a Tactical Cloud, B depicts a Tactical Denial of Sustainability situation (abbreviated as TDoS) if TCS(A,B), and if all the following conditions are satisfied: $C_A \ll C_B$ and $C_A(t) \nsim C_B(t)$; or $tD_A \ll tD_B$ and $tD_A(t) \nsim tD_B(t)$; where:
\end{theorem}

\begin{itemize}
\item $nC$: total cost of upkeeping a particular digital tactical capability.
\item $nC(t)$: distribution of. the cost of upkeeping a particular digital tactical capability as a mission progress.
\item $tD$: total number of tactical nodes that supply certain digital tactical capability.
\item $tD(t)$: distribution of the number of tactical nodes that supply certain digital tactical capability as a mission progress.
\end{itemize}

In these grounds, the conducted research hypothesized about at least a pair of modus operandi able to trigger a potential TDoS situations, which rely on threatening the hybrid upkeep costs (economical, energy, materials, etc.) and/or and the distribution of the digital tactical capabilities thorough the tactical edge. The are defined as Maintenance-based Tactical Denial of Sustainability (U-TDoS) and Deployment-based Tactical Denial of Sustainability (D-TDoS). 

\begin{theorem} \textit{Maintenance-based Tactical Denial of Sustainability}
\label{lemma-3}
Let the state A that depicts the normal and expected behavior of a capability delivered by services enabled by a Tactical Cloud, B depicts a Maintenance-based Tactical Denial of Sustainability situation (abbreviated as U-TDoS) if TCS(A,B), and if the following conditions are satisfied: $C_A \ll C_B$ and $C_A(t)\nsim C_B(t)$. This can be read as that under conventional provisioning conditions (i.e. there are not outlying observations in terms of the ongoing/planned actions nor the number of tactical capabilities that depends on digital tactical capabilities), the upkeep of a digital tactical service became significatively ($C_A\ll C_B$) and unexpectively ($C_A(t)\nsim C_B(t)$) costly at some point of the supported missions.
\end{theorem}

\begin{theorem}  \textit{Deployment-based Tactical Denial of Sustainability}
\label{lemma-4}
Let the state A that depicts the normal and expected behavior of a capability delivered by services enabled by a Tactical Cloud, B depicts a Deployment-based Tactical Denial of Sustainability situation (abbreviated as D-TDoS) if TCS(A,B), and if the following conditions are satisfied: $tD_A \ll tD_B and tD_A(t) \nsim tD_B(t)$. This can be read as that under conventional provisioning conditions (i.e. there are not outlying observations in terms of the ongoing/planned actions nor the number of tactical capabilities that depends on digital tactical capabilities), the tactical nodes that supply certain digital tactical capability somehow increase significatively ($tD_A \ll tD_B$) and unexpectedly ($tD_A(t)\nsim tD_B(t)$) at some point of the supported missions.
\end{theorem}

\subsection{Causality and attack vectors}
Both U-TDoS and D-TDoS may by triggered by multiple and heterogeneous potential threat vectors, which can be linked with vertical (technical, operational) and/or horizontal (tactical) decisions or hostile actions. The technical reasons rely on the vertical propagation from cyber-physical treats to the tactical plane, which should be discoverable by mission-centric risk identification and assessment capabilities \cite{tdos15}, and against responding presents a high dependency of the acquisition of a mission-centric cyber situational awareness \cite{tdos16}. In this context, it is expected that the conventional threat situations on Cloud Computing environments deployed at tactical edge may bring TDoS scenarios, among them  Denial of Service (DoS), Advanced Persistent Threats (APTs), Insiders , etc. \cite{tdos17}\cite{tdos18} The EDoS attacks have the potential of became the key enabler of U-TDoS and D-TDoS, which as demonstrated in \cite{tdos7}, may be enforced by combining some of the aforementioned malicious actions.
The propagations vertical from the operational plane are directly related with the decision-made as the mission evolves, provisioning (e.g. sources of digital tactical capabilities to be onboarded, and the interference of operational-to-tactical C2 communication \cite{tdos19}. For example, the capabilities that enable operational assessment determines the progress of a joint force toward mission accomplishment. An adversarial counterintelligence action may lead to redefine and develop operational plans that place the Tactical Cloud infrastructure in bad terrains, where the communication cost between tactical nodes rapidly drain the energy on which the ICT assets depend. For example, in terms of U-TDoS this may be directly caused by the need for greater energy consumption in order to enable longer-distance communications. On the other hand, D-TDoS may be the consequence of deploying redundant capabilities. The latter may be the cause of Operational-to-Tactical vertical propagations, as well as the consequence of tactical situations. For example, in order to trigger TDoS the adversarial effectors may tactically force to dynamically redistribute the mobile ally tactical nodes, thus leading to a situational similar to that enforced at operational level. In either cases (vertical and horizontal propagations), the denial of the sustainability of the Tactical Cloud have the potential of become a critical tactical problems, on which digital tactical capabilities like C2, ISR or tools for acquiring situational awarenessdepends.
It is important to highlight that the TDoS causality and attack vectors may sinificantly vary depending on the stage of the digital tactical capabilities lifecycle. For example, the attack vectors targeting the onboarding stage are mostly related with the provisioning of the capability dependencies, from which mitigation to protect the end-to-end digital supply chain is critical. In the opposite, attempting to weaponize the adaptation or undeployment stages are mostly related with contextual operational, tactical and technological decision, where the adversarial may attempt to force that the rival enforce particular CoAs in order to gain tactical terrains o made the rival weak them.

\subsection{Impact Dimensions}
Unlike most tactical threats, a TDoS situations have the potential of impacting on multiple and hybrid dimensions, highlighting among them the mission economics, safety, security, energy efficiency or socio-technical features. The economic impacts are related with the software licenses, the scalability/expandability of the digital tactical capabilities, personnel costs, materials, etc. which are expected to pursue deterring purposes \cite{tdos20} as the missions supported by the jeopardized Tactical Cloud extends over time. Alternatively, economic and material expenditures may lead to the prioritization of tactical services, thus exposing new attack surfaces and tactical blind spots \cite{tdos21}. They can derive on new safety and security situations that the adversarial may weaponized towards gain tactical terrains.
The environmental and energy related consequences of TDoS situations on Tactical Clouds can be extrapolated with the economic/material aspects of the mission, which extends towards side-channel related consequences that may facilitate adversarial ISR and Electronic Warfare actions \cite{tdos22}. They may also enforce tactical CoAs like in-flight refuelling, battery recharges, etc. on the ITC infrastructure and its carrying units/facilities, causing delays, exposing effectors and disrupting the original planning of the military missions. 
Finally, the socio-technical impact is related with the Tactical Cloud ergonomics and the human understanding of the Common Operational Picture (COP) \cite{tdos23}. For example, by forcing that the rival deploys a huge number of heterogeneous capabilities at different levels of the tactical environment may be possible to make understanding and prioritization difficult when making decisions, which may be perceived as a socio-technological jamming action. Another more complicated TDoS action against socio-technical dimensions is to enforce that the adversarial deploys digital tactical services that require higher-levels of human capacitation. This would cause that more specific, and hard-to-find profiles lose the focus on other tactical situations.

\section {TDoS against the human sustainability of tactical clouds}
As pointed out by the European Union Institute for Security Studies \cite{newTDoS52}, ''\textit{the digitalization of the military sector should not be seen as silver bullet for every problem facing Europe’s militaries and a human dimension will be required for politico-strategic guidance and maintaining the morale of troops}”. It is inherent to cross-dimensional issues and challenges where novel contributions are not exempt of potentially disruptive changes at all the levels of the Joint Capabilities Integration and Development System (JCIDS): doctrine, organization, training, materiel, leadership and education, personnel, facilities, etc. These changes are directly linked with implications on the human barriers on the acceptance and adoption of the military tactical clouds, all of them needed to be analyzed and considered at design, planning and enforcement of joint operations on the cyberspace. Some of them are enumerated below:

\textbf{Lack of trust to share capabilities and information between military and civilian actors}. The capability of acquiring military sovereign in cyberspace is controversial, since the exclusivity of classical state sovereignty runs contrary to the spirit of the internet, which rests on the concept of unrestricted interconnectivity and net neutrality. On the other hand, most of the required infrastructure and services is managed and/or belongs to the private sector, which places a variety of legal, regulatory, and accepted self-limiting obstacles, hindering public/private cooperation in cyber defense and counterattack \cite{newTDoS53}. Therefore, the development of collaborative platforms to facilitate the information exchange between stakeholders plays a key role in the pursuit of an aligned, connected and prepared cyber defence environment. However, the lack of trust between different public and private actors in each country is seen as the primary inhibitor to cross-sectorial and cross-border collaboration, and intense competition and distrust from business rivals often prevent information exchange and cooperation between different private sector stakeholders \cite{newTDoS54}. Companies and organizations are hesitant to share capabilities with defence stakeholders because of their law enforcement and supervisory functions, as well as fear of sanctions under national laws or competition. In contrast, fear of adverse media coverage is another reason why organizations are reluctant to expose operational information to the public sector and the general public. A clear example of lack of trust and civil acceptation on this cooperation is illustrated in the repudiation of a set of employees, customers and investors of MIT Media Lab with the United States Department of Defence on Project Maven, where AI capabilities were attempted to be developed for supporting combat drone operations \cite{newTDoS55}.

\textbf{Country-specific \& cultural differences}. Notions of privacy, security and safety depend mainly on context and legislation and differ between cultures and countries, even within Europe The last decade has also seen the emergence of novel procedures and concepts of operations with a clear country-specific vision, as is the case of  the regulation and enforcement of ''active cyberdefence” tactics. Lacking a commonly accepted definition, a number of states (e.g., the USA, the UK), international organisations, such as the Cooperative Cyber Defence Centre of Excellence, and scholars have provided explanations that help qualify \cite{newTDoS56}, but they are still far from having a common perception. On the other hand, questions such as infrastructure, security, policy on the transmission of data and sensitive information, awareness and education are not only technical but also, in no small extent, socio-economic and political; linked to JCDP interpretation and enforcement. Finally, it is worth to highlight that many of the national regulations which concern cybersecurity are designed in a similar fashion to regulations relating to public safety and security. Kharlamov and Pogrebna \cite{newTDoS57} hypothesized that a human values-based framework for cybersecurity governance (also applied to Civil-Military Cooperation (CMC)) shall at least ask the following issues: the problem of coexistence between an individual and a group (\textit{society}), coexistence between an individual and social fabric responsibilities (\textit{responsibility}) and coexistence between human beings and nature (\textit{nature}); these answers inherently linked to cultural/regional singularities and heritages.

\textbf{Social perception of digitalization and AI as threats}. Although the digitalization of the defence sector is bringing proven advantages in terms of sustainability, efficiency, quality and performance, it can be perceived by society as a threat, which is directly transposed to the tactical capabilities relying on it. From the workforce standpoint, the main concern is technological barriers and the risk of job loss, but citizens’ point of view, one of the main problems will be the lack of privacy and the risk of continuous surveillance. There is a cross-sectorial resistance to change. Other barriers to social acceptance stem from the emergence of security problems attributed to AI-related processes, which can range from the handling of sensitive information to the presence of cognitive biases in decision stages \cite{newTDoS60} This is the case of \cite{newTDoS61} and \cite{newTDoS62}, which analysed revealed operational failures, partly due to the incorrect use of automatic response systems. On the other hand, in \cite{newTDoS63} the risk and  (un)intended of intelligence agencies behaviours in cyberspace are reviewed in detail, among others concluding that there is a need of exploring possibilities to define objects and organisations that should be off limits for cyber operations, in line with the ‘Geneva’ style of the regulation of responsible state behaviour; also suggesting to resist the temptation of replying in kind to influence operations that target the integrity of public information, since integrity of data and information – much more than confidentiality and availability – touches on core values of democratic societies and resembles a very negative social perception.

\textbf{Misuse and other human factors}. Morality and ethics have occasionally been identified as precursors to or inhibitors of important phenomena in the behavioral security domain, which in absence may derive in insider threat triggering situations categorized as traitors and masqueraders, the first (traitors) resembling legitimate workforce trying to gain privileges for accessing restricted assets/capabilities with malicious purpose, and the second (masqueraders) entails external actors that are somehow able to impersonate authorized users and disrupt planed/ongoing tactical operations \cite{tdos17,tdos18}. Both scenarios are catalyzed by the lack of training, motivation, understanding and awareness of the personnel; which are triggered by assuming the gaps and challenges described above. On the other hand, human factors may mislead operations relying on digitalization and tactical clouds inherent in the human cognition and capability to react to a massive digest of information. In this regard, Research on Human Cybersecurity (HCS) behavior suggests that for example, time pressure is one of the important driving factors behind non-secure decisions \cite{newTDoS64}. The research on facilitating the acquisition of cybersituational awareness also revealed the need for capabilities able to adapt the presentation of cross-domain operational pictures to the decision-making contexts, so human beings behind decisions can efficiently consider the most relevant information \cite{newTDoS65}.  

\subsection{Enterprise, Organizational and Individual TDoS threats}
In 2015, Gen. Odierno commissioned the U.S. Army Human Dimension Strategy (AHDS) that outlined the way ahead for development of human dimension assuming a time horizon of ten years \cite{newTDoS66}. The paper defined the human dimension as the cognitive, physical, and social components of the Army’s trusted professionals and teams, emphasizing the need to ``\textit{optimize the human performance of every soldier}” and ``\textit{build cohesive teams of trusted professionals who thrive in ambiguity and chaos}”. As response, the strategy defined the US Cognitive Dominance Line of Effort (LOE) concerning objectives and tasks that equip Army personnel with the intellectual aptitude, cultural understanding, physical toughness, and resilience to adapt and thrive in ambiguity and chaos. The conceptualization of the TDoS on the cognitive dominance can be abstracted according to the model describe by C. Song \cite{newTDoS67}, which by exploring the transformation of armies for attract retain and develop both civilian and military personnel, perceived model armies as ``systems,” being larger military forces as ``System of Systems” (SoS); i.e. ``\textit{assemblages of components which individually may be regarded as systems, and which possess two additional properties - operational independence of the components and managerial independence of the components}”. Accordingly, the human impact derived from adversarial activities on the tactical edge can be framed as \textit{enterprise}, \textit{organizational} and \textit{individual} levels, which are enunciated at the following lemmas.

\begin{theorem} \textit{Enterprise-based Tactical Denial of Sustainability}\\
\label{lemma-1}
Let the state $A$ that depicts the normal and expected behavior of a capability delivered by services enabled by a Tactical Cloud, $B$ depicts a Enterprise-based Tactical Denial of Sustainability situation (abbreviated as E-TDoS) if jeopardizes the army human capital lifecycle management system concerning recruiting, trusting, training, educating, developing, promoting and/or retaining both military and civilian talent \cite{newTDoS68}.
\end{theorem}

\begin{theorem} \textit{Organizational-based Tactical Denial of Sustainability}\\
\label{lemma-1}
Let the state $A$ that depicts the normal and expected behavior of a capability delivered by services enabled by a Tactical Cloud, $B$ depicts a Organizational-based Tactical Denial of Sustainability situation (abbreviated as O-TDoS) if jeopardizes the effectiveness, efficiency and upkeep of governance and business practices aimed on accelerating communication, decision-making, and DOTMLPF-P integration for military human resource management.
\end{theorem}

\begin{theorem} \textit{Individual-based Tactical Denial of Sustainability}\\
\label{lemma-1}
Let the state $A$ that depicts the normal and expected behavior of a capability delivered by services enabled by a Tactical Cloud, $B$ depicts a Individual-based Tactical Denial of Sustainability situation (abbreviated as I-TDoS) if jeopardizes the independent actions of each individual within the army. The AHDS organizeds the individual situations into three core groups, which may overlap: social, cognitive and physical situations \cite{newTDoS66}.
\end{theorem}

\subsection {Socio-technical dimensions and attack vectors of TDoS}
The adversarial actions against the military enterprise (E-TDoS) and organizational (O-TDoS) by targeting the tactical cloud inherently entail vertical propagation from tactical to operational and strategical goals thorough exploiting cross-frontier vulnerabilities across the Political, Military, Economic, Social, Informational and Infrastructure (PMESII) spectrum \cite{newTDoS69}. Consequently, they are connected with hybrid warfare situations where the cyberspace operates as a highway interface that links the ``grey zone “ with cybernetic assets deployed at the tactical environments on which the tactical cloud relay on, or which are enabled by this interface; and operational/strategical capabilities vertically connected with their related technical actions. Therefore, tactical clouds enable network-centric warfare tactics able to trigger hybrid consequences \cite{newTDoS78}. Some of the most probable nonviolent hybrid threat instruments inferable from digital tactical capabilities are reviewed in \cite{newTDoS70}, as is the case of services linked to media/propaganda, military intelligence, communication and networking, or economical sustainability. By making their upkeep more expensive, forcing the fraudulent instantion of poisoned related VNFs or sinkholing targeted groups of tactical services, the conventional U-TDoS and D-TDoS effects derived from cyber attacks like I-EDoS and W-EDoS have the potential of impact on the military enterprise and organization from their bases (comprimising due to cognitive and physical factors their recruitment effectiveness, trustworthiness, poisoning of the aquired institutional picture of the theaters of operations, etc.). The U.S. Institute for the Study of War highlighted a subset of behavioral patterns for grey zone operations that can be directly transposed to the enforcement on E-TDoS and O-TDoS: 1) creation of maximum uncertainty at the tactical edge: covert and clandestine actions preferred, 2) maintenance of  deniability of E-TDoS and O-TDoS as long as possible, 3) gradually escalating pressure, indirectly triggering ``domestic” violence as necessary, 4) staying below the threshold of war or triggers of intervention (at least while E-TDoS and O-TDoS aim to proxy/hybrid goals), 5) when aiming to trigger direct human reactions, mixing “carrots” and ``sticks , and 6) last resort or when ``invited” conventional military force. Bearing all this in mind, the close dependencies between cyberspace and information warfare makes the tactical clouds perfect platforms for deploying related defensive effectors, in the same time that expose a complex potential attack surface \cite{newTDoS72}.

The hybrid warfare and the use of cyber assets as part of it is one of the most important factors for understanding the future arc of conflict \cite{newTDoS73}, which beyond jeopardizing the socio-technical dimensions of the military enterprise and organization layer, may directly impact on the individual at any level (including technological and tactical) resulting in I-TDoS situations. A good example of the adoption of cyber effectors against socio-cognitive aspects was observed during the Russo-Ukrainian conflict at 2014, where combat actions in Illovaysk and Debalcevo were preceded by a significant burst of activity in information space. These theaters of operations identified negative information on key authorities of Armed Forces of Ukraine and government representatives as a cyber aggression, coupled with disinformation from proxies and false fronts on the internet. In related  contexts, a tactical cloud resemble a perfect vector for deploying disinformation actions while jeopardizing the required ICT capabilities towards maximizing their dissemination effect. These enablers were discussed in deep in \cite{newTDoS73}, highlighting anonymous claims to authority, news items manipulated with half-truths, repetition of messages, information overload, cyber-pseudo operations (government posing as insurgents), sock-puppeting (government agents playing the role of online commentators), and astro-turfing (creating of false grassroots movements).

Finally, I-TDoS attacks triggered from tactical clouds or against them may take advantage of the data processing capabilities bring by the digitalization of the battlefield for disrupting the human ability of understanding the operational picture; especially if this is facilitated by AI-driven automatism. For example, from \cite{newTDoS75} can be inferred that data dependent AI enablers may be manipulated or poisoned for affecting the human decision-making activities that are feed by them; TDoS actions may lead to similar results by overload of targeted edge services and leading the victim to develop a biased perception of the operational situation. Related risks inherent in data forgery and adversarial AI tactics when applying federated machine learning at the tactical edge are described in \cite{newTDoS76}, where it is emphasized that tactical cloud services may face rapidly changing, never-before-seen situations, where pre-existing training data will quickly become ineffective (e.g. tactical training data is noisy, incomplete, erroneous, etc.); and exposes adversarial opportunities for enforcing deception tactics on the collected information required for it update/upgrade \cite{newTDoS77}. From the TDoS perspective, this may occur among others by forcing the fraudulent instantiation of VNFs from poisoned VF inventories (e.g. third-party digital marketplaces) thus jeopardizing the data processing or information exchange processes; or by injecting via I-EDoS a false perception of activity able to create and/or exploit blind spots in the automatism. As more extended examples, the cases of study presented as CONOPs in Section 7 hypothesize on additional vectors against the human sustainability framed in the concrete mission narratives that they describe.

\section {Jeopardizing the energy efficiency of tactical clouds}
As indicated by Samaras et al, ``\textit{Energy has played a role in every facet of war from troops in garrison and defensive planning to mobilization and attack. The need to deliver adequate and timely energy supplies to military forces—particularly to those in the most forward-deployed locations—has long existed as a strategic vulnerability to the success of military campaigns}” \cite{newTDoS88}. Energy resembles a critical aspect to consider from classical and modern defense planning; and at all warfare levels (strategical, operational, tactical and technological). The digitalization of the sector and management of tactical clouds is not exempt of these considerations; in fact, the instantiation, upkeep and removal of digital tactical capabilities typically demands a movement of high volume of data across different geographically separated nodes which create a huge burden on the network infrastructure with strong energy and environmental implications \cite{newTDoS79}. Since to extended the tactical cloud lifetime with complete connectivity and coverage entails a critical edge terrain \cite{newTDoS82}, ongoing and ambitions defense projects like the DARPA’s Near Zero Power RF and Sensor Operations (N-ZERO) \cite{newTDoS81} address the challenges of extending the energy efficiency, coverage and connectivity of related enablers (cyber-physical sensors, wireless communication or distribute data storage capabilities, etc.). In this context, Wang et al. [80] pointed out three core actions towards minimizing the energetic and environmental impacts derived from tactical digital information processing: 1) to enforce effective workload classification so that ``latency-hungry” and realistic end user requests can be provisioned to edge devices and ``resource-hungry” requests are relayed towards central computing capabilities (e.g. data centers); 2) to optimal scheduling the classified workload in edge–cloud environment in order to achieve trade-off between energy efficiency and effectiveness; and 3) to migrate tactical requests to centralized computing capabilities when the digital tactical capabilities are enabled by resource-limited edge devices. 
But despite the challenges inherent in the deployment, supply and maintenance of digital tactical services, adversaries may aggravate related situations by enforcing hostile actions against the tactical cloud lifecycle and energetic dependencies. Threats like the depletion-of-battery attacks \cite{newTDoS83}, wormhole attacks \cite{newTDoS84}, clone attack (nodes replication attack) \cite{newTDoS85}, vampire attacks \cite{newTDoS86}, or denial-of-sleep attack for keeping the sensor nodes awake to consume more energy \cite{newTDoS87};  demonstrated effectiveness when causing energy depletion at dual use data processing services. 

As proven in \cite{tdos7}, the EDoS threats that target the instantiation (I-EDoS) and workload (W-EDoS) of virtualized serviced deployable at the edge can effectively cause cost overrun linked to fraudulent CPU, memory, connectivity, bandwidth, etc. usage; all of them transversely derived in malicious energy expenditures. Based on them, specific hostile actions against digital tactical services can be driven by the M-TDoS and D-TDoS actions; thus compromising any phase of the tactical services provisioning and upkeep (onboarding, deployment, surveillance, adaptation, etc.). It is important to consider that the Allied Joint Doctrine for the Planning of Operations explicitly refers to the critically of the sustainment concept as: ``\textit{No COA is complete without a plan to sustain it properly. The sustainment concept is more than just gathering information on various logistic and personnel services.}”. In this context, the Allied Joint Doctrine for the Conduct of Operations \cite{newTDoS90} points out success conditions dependant on the energetic sustainability of the digital tactical services, among them the Ally unity of effort, concentration of force, flexibility, surprise, agility, or synchronization. Any of the aforementioned technical-level threats against the energetic sustainability of the cloud can be vertically propagated to tactical and operational situations and dimensions, some examples of this being illustrated in the CONOPs presented in Section 7.

\section {TDoS in Military Thinking at the cyber Tactical Edge}
Kallberg and Cook \cite{newTDoS91} pointed out four core premises to be assumed when conducting cyber military operations: 1) lack of object permanence, which undermines the concept of maneuver; 2) limited or absent measurement of effectiveness in offensive cyber; 3) conflicts that are executed at computational speed, removing the time window that would allow for meaningful strategic leadership; 4) anonymity that makes the parties to the conflict unsure who is the other party. These assumptions make the conventional C2 thinking directly linked to OODA (Observe, Orient, Decide, Act) loop weaken, since they compromise their four tenants, and thus project the need for significant changes at doctrine-level as the defene sectors is digitalized. Complementarily, and as remarked by Schulze \cite{newTDoS93}, the ``\textit{tactical cyber operations are difficult to integrate into the traditional target cycle of conventional forces due to their long planning and development time. Traditional weapons only need to be targeted once; tactical cyber operations must provide permanent covert access to a hacked system}”. Because they need significant preparatory and enforcement time, nowadays cyber actions are mostly perceived as strategic decisions. From the tactical level, the fact that they require a constant spatial proximity with the adversarial CIS infrastructure make their execution under traditional circumstances very challenging; and this is exactly the point where the tactical cloud and digital tactical capabilities entails a promising enabler. Since the research presented in this paper focused in the tactical edge and its digital tactical capabilities, it is possible to assume that these conditions will also impact on the tactical operations at the cyber edge. In this context, TDoS should be considered as potential effectors operable at offensive and defensive doctrine-level. In order to contribute to their adoption, as well as for the better understanding of their potential, the following introduces some key considerations on their application at offensive and defensive thinking processes.

\subsection{TDoS in Offensive Thinking}
The historic of military thinking has been mostly approached by different ways and variations of offensive thinking, most of them inherited from the widely adoption of OODA loop based C2 processes; as OODA is driven by the idea of getting inside the enemy’s decision cycle \cite{newTDoS92}. This loops has successfully applied defensively, but requiring to observe the attacker actions and with the requirement of disposing the capability of fast orient, decide and taking (counter) actions, in this way presenting an effective prevention model. But as discussed in \cite{newTDoS91}, TDoS operations and the nature of the cyberspace rely on assumptions that suggest the need for extending the conventional offensive models, which shall accommodate to the doctrine on Offensive Cyberspace Operations (OCO) \cite{newTDoS94}. Accordingly, TDoS attack capabilities shall create fires in and thorough cyberspace, occasionally leading to desired physical alterations, the latter potentially deriving in cascading effects (including collateral effects) in the rest of kinetic warfare domains. These actions shall encompass a number of task, action and processes (e.g. targeting, coordination, deconfliction) according to the enforced fire doctrine \cite{newTDoS95}, typically aiming on jeopardizing the sustainability of adversary digital tactical capabilities, or triggering first-order sustainability effects to initiate carefully controlled cascading situations at strategical, operational or tactical levels. 

TDoS fires may deny (degradate, disrupt, destroy) or manipulate controls or information towards cause deception, decoying, conditioning, spoofing, falsification, and other similar effects. Layton \cite{tdos41} discussed four different illustrative approaches for offensive thinking applications from tactical clouds at 5th air warfare operations, which can easily benefit from TDoS effectors: 1) \textit{Attack Adversary Sensing Grids}: offensive TDoS can contribute to reduce their autonomy, degrade their capability of collecting and managing information, manipulate their observations, and from the latter, trigger cascade hybrid impacts (loss of confidence, propaganda, jeopardizing supply chains, injection of noise for diffusing the information from which the operator constructs the operational image, etc.). 2) \textit{Attack an Adversary’s Long-Range Strike Systems}: Layton conceptualized the strike systems as an adversary grid of short-range nodes but considerably fewer long-range strike nodes. Accordingly, long-rage nodes became a practical target set for friendly forces to engage and achieve meaningful tactical results, which infrastructure and CIS level assets can be targeted by each of the offensive actions described above. 

Layton also remarked that long-range strike are more expensive and difficult to replace, typically depending of more complex monitoring and analytical capabilities. 3) \textit{Rapidly Attrite Adversary Forces}: This is one of the most challenging scenario discussed by Layton , where higher loses may be affected in order to gain a fast victory, typically by making simultaneous attacks converge as parallel air warfare. Significant intelligence on the adversarial tactical cloud is needed, and cyber effectors shall be able to appropriately scale towards absorb combat losses to continue with the tempo. This scenario will mostly require TDoS capabilities able to deplete the resources of the attacked cloud and thus forcing tactical decisions (U-TDoS, I-TDoS) while feeding information able to create confusion and disguise the rest of the convergence actions (I-TDoS, O-TDoS, etc.). Finally, Layton discussed the 4) \textit{Force Horizontal Escalation} offensive scenario, which is the closer to the TDoS concept. In this context, TDoS may force adversarial capabilities do distribute over the theater of operations  and/or move from specific terrains, so allies can exploit the triggered situation for reorient their efforts and address high-value regions. Socio-technical TDoS actions and attacks reducing the autonomy, energy or digital resources of the tactical cloud and/or its sensors grids entail promising support for these actions. More detailed examples of TDoS offensive thinking against tactical clouds can be found in the Concepts of Operation presented at the next section.

\subsection{TDoS in Defensive Thinking}
The adoption of a defensive thinking posture may be subject to different conditions, including military superiority, tactical restrictions or strict Rules of Engagement (ROE). As occurs with conventional cyber situations, attackers have the advantage of surprise, and plenty of time to gather the cyber threat intelligence needed to success on the TDoS actions (enumeration of attack surfaces, discovery of vulnerabilities, customization of exploits, etc.) \cite{newTDoS97}. On the other hands, defenders tend to decide the cyber operational context (hardening, decoys, passive security procedures, etc.) and can recall intelligent on the attacker modus operandi \cite{tdos41}. As defined at the UK AJP-5 \cite{newTDoS89} “\textit{in defensive operations, the defending force reaches its culminating point when it no longer has the capability to mount a counter offensive or defend successfully and needs to be reinforced, disengaged or withdrawn to avoid defeat}”. Transposed to the tactical cloud concept, the goal is to defeat the threat of a specific adversary and/or return the cloud or the affected digital tactical services to a secure and functional state. 

Similarly to the NATO JP3-20 \cite{newTDoS94}, defensive actions on tactical clouds may entail Internal Defensive Measures (IDM), Defensive Response Actions (DRA), or the combination of both of them at owned or ally cyberspace. The first occur within the defended edge infrastructure and capabilities, are focalized into dynamically reconfirm or reestablish the security of degraded, compromised or threatened assets by rerouting, restoring, isolation, decoying, etc. In \cite{tdos7} passive defensive tactics able to discover and thwart denial of economical sustainability attacks against commercial cloud environment were introduced. Similarly to other state of the art solutions, they focused on prevent the fraudulent scaling of the computational resources expenditures by identifying ``lazy and unproductive” virtual machines while isolating those that may contribute to the horizontal propagation of the threats. Although it is out of the scope of this paper to propose and discuss potential anti TDOs procedures and algorithmic, it has sense to hypothesize that related solutions can apply against the U-TDoS and D-TDoS modus operandi introduced at Section 3. In particular, U-TDoS defense shall focalize on preventing and detecting fraudulent and unexpected maintenance issues on the deployed digital tactical capabilities; while D-TDoS mitigation may address the optimization of the digital tactical service instantiation and the dynamic removal/recovery of those fraudulent instantiated.

On the other hand, DRAs actions are characterized by taking place out of the defended portion of the edge, and without authorization of the owner or responsible of the external assets. These defensive procedures tends to require very precise objectives, strict ROEs and a proper justification, since they may be confused with cyber offensive actions or settled out of \textit{just ad bellum}. Active defence against TDoS  shall focalize on the sources of the hostile activities (which may be a network, hostile digital tactical capabilities, etc.) and block, trace, neutralize, etc. the enemy physical or digital assets linked to the TDoS attack affection. General purpose active response procedures may serve for this aim, while more specific actions may explore SON enablers towards orchestrate quarantine regions, sinkholes, etc. guided from isolate the source regions within the external CIS edge. Active TDoS may attempt to scale to socio-technical aspect at tactical level (I-TDoS) or vertically propagating up to organizational or strategic levels (O-TDoS, E-TDoS)

\begin{figure*}[t]
  \centering
  \includegraphics[width=0.6\linewidth]{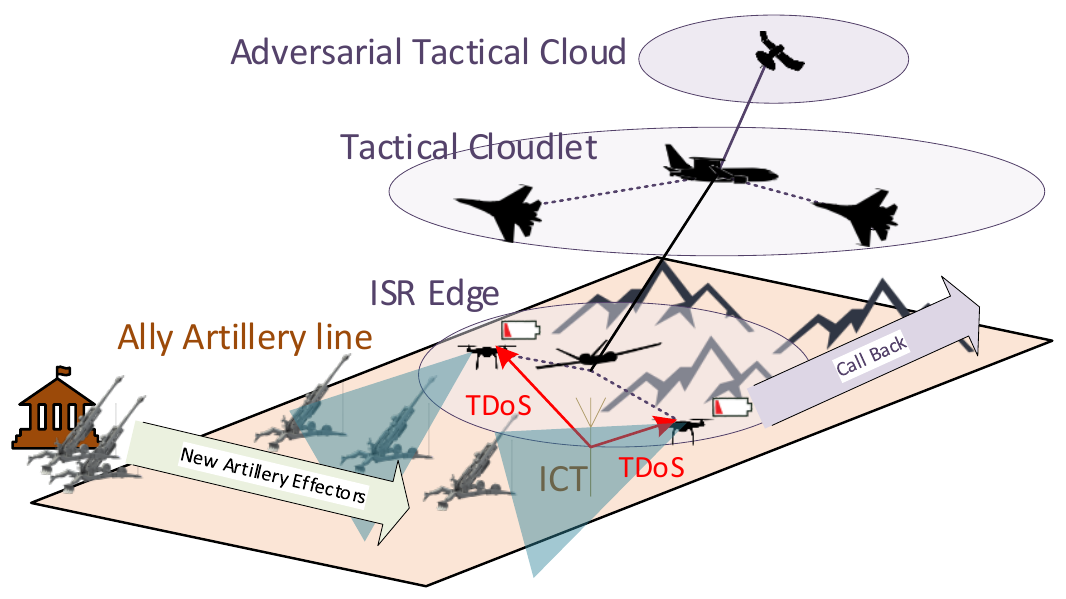}
  \caption{Hybrid TDoS fires agaisnt adversarial ISR operations}
  \label{fig2}
\end{figure*}

\section{Cases of study}
In the grounds of the COPD doctrine \cite{tdos24}, this sections embraces the purpose of defining illustrative Concepts of Operation (CONOPs) for  hypothesized tactical offensive/defensive capabilities able to develop TDoS situations on adversarial Combat Clouds. The concepts explore to jeopardize key elements of fusion warfare and the Joint Aerial Layer Network (JALN) \cite{tdos28} linked to C3 services critical to the proper achievement of tactical goals. In order to facilitate the understanding to the Tactical Cloud applications and their direct dependency with the sustainability of their digital tactical capabilities, different scenarios have been developed assuming hybrid warfare \cite{tdos25}, proxy warfare \cite{tdos26} and asymmetrical warfare \cite{tdos27} situations.

As remarked in Section 1, it is important to highlight that the introduced cases of study are fictitious scenarios with a mere didactic purpose, where  most of the assets and processes inherent in related military operations have been omitted for simplicity and in the shake of learning, only describing the technological and procedural aspects directly linked to the TDoS situations. As in the previous sections of the project, any deductible Rules Of Engagement (ROE), \textit{jus ad bellum}, or \textit{jus in bello} premise do not intentionally align with any particular international agreement. Factions, terrains and actors described do not correspond in any way to real entities.

\subsection{CONOP\#1: Jeopardizing the adversarial energy efficiency at ISR operations}
\textbf{Background:} A peace agreement between two confronted nations has resulted in a fragile ceasefire. The ally faction strategically decided to take the opportunity for secretly rearming reinforce their anti-aircraft artillery systems with a novel target development system. The ally intelligence reported that the adversarial faction is conducted ISR actions in the surrounding of the anti-craft effectors supported by a Tactical Cloud.

\noindent \textbf{Current situation:} The confronted factions have a common military doctrine. In both cases the humanitarian situation is approaching crisis proportions. 
\begin{itemize}
\item \textbf{Own Forces}. The ally forces are less technological advanced. The adversarial reached air, space, and sea superiority, but ally forces have land superiority. 
\item \textbf{Co-operating and neutral actors}. Civilian supporters may cooperate in easy surveillance and logistic operations.
\item \textbf{Assumptions}. 1) Conventional adversarial air ISR operations are not capable of discovering the new anti-craft effectors. However, certain digital tactical capabilities may reveal these assets if properly deployed at the edge by a Tactical Cloud. 2) The hybrid computing services that enable the Tactical Cloud are vulnerable against certain W-EDoS attacks, which can be enforced from conventional ITC infrastructure. 3) The mobile tactical nodes are feed by short/mid duration electronic batteries, which energetic consumption varies on the workload of the C3 services they enable. 
\item \textbf{Limitations and constraints}. The ally forces have not a Tactical Cloud nor physical way to disrupt the operations of the adversarial digital tactical nodes. 
\item \textbf{Tactical end-state and objectives}. The end-state is achieved when the anti-aircraft effectors are properly rearmed. The tactical objective of the operation is to prevent that de digital tactical services discover the rearming operations.
\end{itemize}

\noindent \textbf{Mission:} The digital tactical capabilities able to discover the rearming anti-aircraft artillery shall be denied or delayed enough time for properly completing the rearming actions without being notified by the adversarial ISR actions (see Fig. \ref{fig2}).

\noindent \textbf{Execution:} The commander purpose is to delay the activities of the digital tactical capabilities as long as needed for rearming of the anti-aircraft artillery systems without being discovered by the adversarial. With this purpose, the commander will conduct U-TDoS by firing W-EDoS attacks to the technological infrastructure that supports the adversarial digital tactical capabilities. Their overload shall increase their energetic dependence, thus forcing the need for more frequent recharge and battery provisioning, which will delay the tactical activities the enable.

\noindent \textbf{Tasks:} 1) Enumeration and exploitation of the cyber attack surfaces of the CIS infrastructure that enables the adversarial tactical nodes. 2) Exploitation of the identified vulnerabilities and fire of W-EDoS cyber attacks against them. 3) Coordination with the command of the anti-aircraft artillery rearming operation towards ensure that the cyber fires are prolonged until the end-state is reached. 4) Remove any evidence about this cooperation, and the offensive modus operandi.

\noindent \textbf{Coordination instructions:} Cyber Threat Intelligence is dynamically gathered and distributed between the participant cyber commands. Information about the status of the rearming actions is shared between their commanders. A collaborative cyber situational awareness is developed, including a join COP. The liaison, engagement rules, communication procedures and reporting systems are those comprised by the allied doctrines.

\noindent \textbf{Integrated support systems:} Cyber fires are commanded from the ally cyber command HQ. 

\subsection{CONOP\#2: TDoS against massive disinformation proxy tactics}
\textbf{Background:} Two factions are in long-term confrontation despite the bad societal perspective of the situation. Consequently, they use to rely on proxy-war tactics like funding, military training, arming, etc. triggering the hostile actions by co-operating states and related stakeholders. In this context, local adversarial militias are called to arms through broadcasted cyber propaganda, which among others is issued by digital tactical capabilities serving a Tactical Cloud deployed on the conflict zone. The propaganda is delivered by spam services able to discover and exploit common vulnerabilities on the Data Link and Network Layers (Ethernet, ZigBee, WLAN, etc.) \cite{tdos29}, injecting them on the civilian technological devices.

\noindent \textbf{Current situation:} The direct confronted factions have a common military doctrine. Their proxy cooperators are in a direct hostile situation that involves both military and civilian actions.

\begin{itemize}
\item \textbf{Own Forces}. The proxy cooperators are in a similar technological capacitation. The superiority of the cyberspace is in dispute, as is the case of that on the rest of the operational domains.  
\item \textbf{Co-operating and neutral actors}. The ally proxy state facilitates its ICT communication infrastructure, which among others includes 5G capabilities connecting large metropolitan areas \cite{tdos30}.
\item \textbf{Assumptions}. 1) The enumeration and intelligence services reported that the adversarial propaganda systems dynamically detect vulnerable end-to-end metropolitan communications, and inject the propaganda as man-in-the-middle actions. 2) The propaganda reached the civilians via unwanted ads, pop-ups and other classical spam actions. 3) The targeted metropolitan areas have large Cloud-Ran (C-RAN) infrastructure with real time network virtualization capabilities \cite{tdos31}. 3) The C-RAN virtualizations horizontal-scale, so during the TDoS actions they shall not suffer computational nor energy restrictions. 4) The adversarial Tactical Cloud nodes have strong computational limitations, which are overcomed by instantiation of on demand digital tactical services thorough the edge.
\item \textbf{Limitations and constraints}. The ally forces have not a Tactical Cloud nor DDoS/jamming capabilities able to fight back or prevent the emission of cyber propaganda.
\item \textbf{Tactical end-state and objectives}. The end-state is achieved when the cyber propaganda stops to reach the civilian population. The mission objectives is to prevent that the emitted cyber propaganda onboard on civilian devices. 
\end{itemize}

\begin{figure*}[t]
  \centering
  \includegraphics[width=0.6\linewidth]{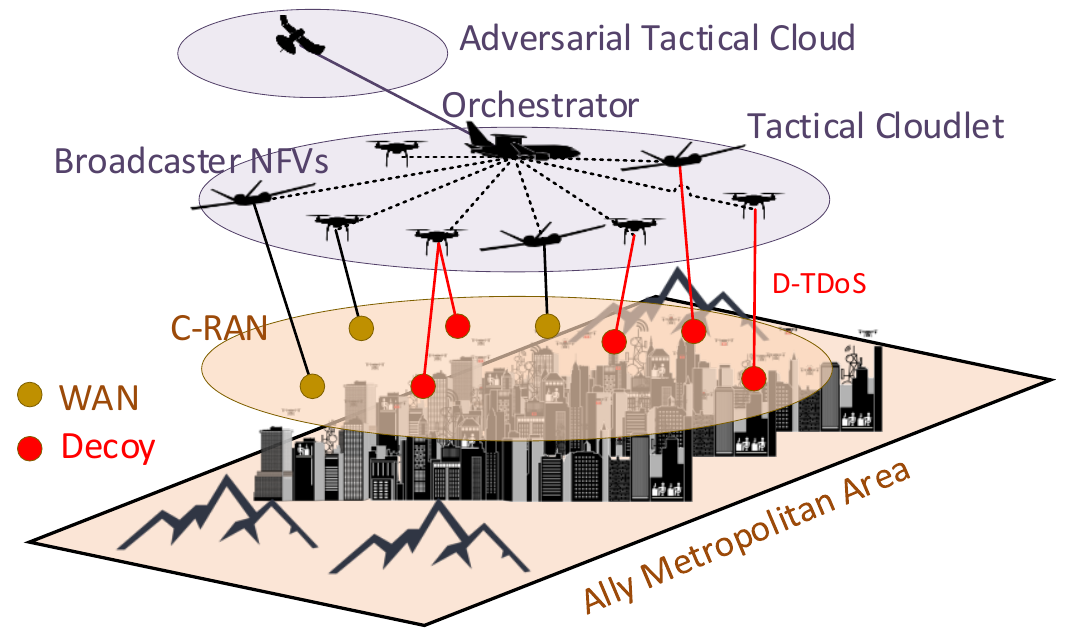}
  \caption{TDoS against massive disinformation proxy tactics}
  \label{fig3}
\end{figure*}

\noindent \textbf{Mission:} The ally cyber commands shall cooperated with the proxy ally state towards weaponing its available 5G infrastructure towards TDoS-based mitigate the propaganda emitted by the adversarial Combat Cloud (see Fig. \ref{fig3}). 

\noindent \textbf{Execution:} The commander purpose is to dissipate the adversarial cyber propaganda before it reaches the civilian devices by exploiting D-TDoS. In particular, the existing C-RAN capabilities of the proxy state shall be able to procedural generate decoy networks and end-point behaviors, so the Tactical Cloud direct their activities against them rather than the real civilians communications. If the number of decoys is large enough 1) most of the propaganda will be redirected to cyber sinkholes, and 2) the Tactical Cloud will need to expand enough for covering most of the faked civilian assets. Since its scalability is limited, this will lead to a D-TDoS situation.

\noindent \textbf{Tasks:} 1) Synchronization with the proxy state cyber commands. 2) Enumeration of the adversarial propaganda actions and the behavior of its enabling Tactical Cloud. 3) Development and fire of decoy  networks, end-users and ICT virtual infrastructure attempting to cause a larger deployment on digital tactical nodes. 4) Sinkhole the propaganda that reach the faked assets. 5) Remove any evidence about this cooperation, and the offensive modus operandi.

\noindent \textbf{Coordination instructions:} Cyber Threat Intelligence is dynamically distributed between the participant cyber commands and the proxy effectors. A collaborative cyber situational awareness is developed, including a join COP.

\noindent \textbf{Integrated support systems:} Cyber fires are supported by the ally cyber command HQ. However they are directly commanded by the proxy state Minister of Defence (MoD).

\subsection{CONOP\#3: TDoS fires supported by air combat operations}
\noindent \textbf{Background:} Two rival factions are engaged in a short term conflict, which included direct and proxy hostile actions. Their military capabilities are symmetrical, and both are able to combine conventional kinetic actions with operations on the cyberspace. Both actions have symmetrical Tactical Cloud capabilities. The ally intelligence services reported that the adversarial is preparing a big drone offensive against critical terrains thorough the conflict area

\noindent \textbf{Current situation:} The direct confronted factions have similar military doctrine.
\begin{itemize}
\item \textbf{Own Forces}. The direct confronted factions have similar technological capacitation and effectors. The ally faction is closed to reach the superiority of the cyberspace. On the opposite, the adversarial faction is closer to reach air superiority. The rest of the operational domains are under balanced dispute. 
\item \textbf{Co-operating and neutral actors}. There are not potential cooperators nor neutral actors. Military engagements are far from civilian actors, typically under BLOS conditions.
\item \textbf{Assumptions}. 1) The foreseen adversarial offensive will be enforced by a swarm of Remotely Piloted Aircraft Systems (RPAS) \cite{tdos32}. 2) The RPAS will rely on digital tactical services provided by a Tactical Cloud, that among others will enable their air-to-air communications and C2 services. 3) Under close combat engagements, the digital tactical services will prioritize enabling real-time communications and C2 actions above cyber defence and/or other cross-cutting services \cite{tdos33}. 4) Due to the BLOS condition, the adversarial Tactical Cloud vulnerabilities are only exploitable by close dynamic tactical nodes, this disabling any potential counteraction from fixed conventional ITC infrastructure. 5) Due to the adversarial air superiority, the ally air forces are only able to delay the adversarial of reaching its goals, which suggests the need for a joint action.
\item \textbf{Limitations and constraints}. The enforcement of jamming and denial of service related CoAs is not recommended by the analysts, since it may jeopardize the ally defence communications, and would not be able to cover the entire space of operations.
\item \textbf{Tactical end-state and objectives}. The end-state is achieved when the adversarial air actions are neutralized or no longer risks the targeted key terrains. The mission objective is to disrupt the foreseen adversarial offensive actions in order to protect their targeted terrains.
\end{itemize}

\begin{figure*}[t]
  \centering
  \includegraphics[width=0.6\linewidth]{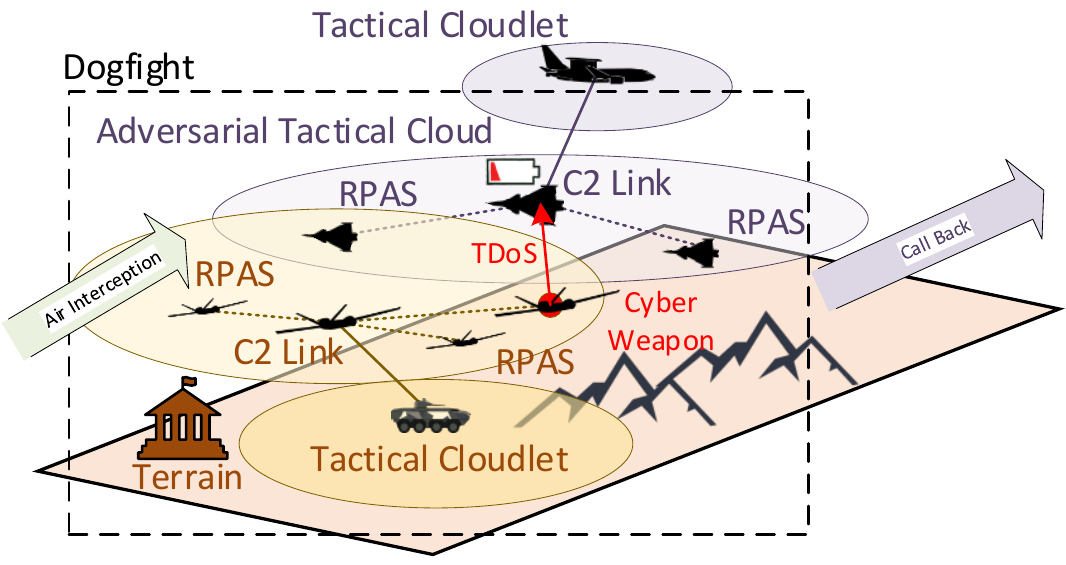}
  \caption{TDoS fires supported by air combat operations}
  \label{fig4}
\end{figure*}

\noindent \textbf{Mission:} A joint operation between the air forces and the cyber commands shall defence the key terrains targeted by the adversarial air operations (Fig. \ref{fig4}).

\noindent \textbf{Execution:} A joint defence between ally air forces and cyber commands shall contain the RPAS threat until a close combat situation exposes exploitable attack surfaces on the adversarial digital tactical capabilities. When this occurs the ally Tactical Cloud will onboard and deploy cyber weapons able to trigger W-EDoS and I-EDoS on the adversarial VNFs orchestration services that enable C2 and air-to-air communications between the RPAS that integrate the hostile swarm. The denial of the adversarial tactical sustainability shall force it to retire. If the tactical deployment of cyber weapons on the ally Tactical Cloud is fast and effective enough, when the enemy retires the ally air forces will have contained the kinetic impact on the protected terrains.

\noindent \textbf{Tasks:} 1) Synchronization between air and cyber effectors. 2) Detection of the RPAS-driven large-scale offensive against key ally terrains. 3) enforcement of air close combat situations in order to press the adversarial Tactical Cloud towards prioritizing Real-time and C2 communication digital tactical services above those related with cyber defense. 4) Enumeration of the cyber vulnerabilities of the adversarial Tactical Cloud. 5) Exploitation of the identified cyber attack surfaces towards causing W-EDoS and I-EDoS. 6) Extend their propagated TDoS situations as much as possible in order to force the adversarial retirement. 7) Removal of evidences of the conducted cyber operations.

\noindent \textbf{Coordination instructions:} A collaborative cyber situational awareness is developed, including a join COP. The liaison, engagement rules, communication procedures and reporting systems are those comprised by the allied doctrines.

\noindent \textbf{Integrated support systems:} Cyber fires are commanded from the ally cyber command HQ, and air actions are commanded from the air command HQ. They are coordinated from a join operaton HQ.

\subsection{CONOP\#4: Jeopardizing the Digital Supply Chain}
\textbf{Background:} At a mid-term conflict between an ally and adversarial faction, human intelligence sources reported an incoming enemy infiltration, which will be preceded by advanced ISR actions driven by never-seen-before hostile digital tactical services enabled by the sensing grid, from an adversarial tactical cloud. The disruptive digital tactical capabilities are no deployable from the current adversarial tactical repository of Virtual Functions (VFs), so they require to be requested and provisioned from a digital supply chain directly connected to a global “distribution market” \cite{newtdos98}.

\noindent \textbf{Current situation:} The confronted factions have a similar military doctrine concerning tactical cyberspace operations.. 
\begin{itemize}
\item \textbf{Own Forces}. The confronted factions have similar technological capabilities and effectors. There is not a clear cyberspace superiority. Ally forces have fixed ICT infrastructure able to reach the adversarial VNF orchestration and management capabilities.  Adversarial digital tactical services coordinated from them are able to reach an external artefact repository (VF marketplaces), from which new digital tactical capabilities can be provisioned to the tactical edge. 
\item \textbf{Co-operating and neutral actors}. . Civilian neutral actor and ICT infrastructure.
\item \textbf{Assumptions}. 1) The adversarial disruptive ISR functions may change the curse of the cyberspace dominance. Once deployed, these capabilities can be reconfigured and adapted to the operational context. 2) Adversarial digital tactical capabilities as VNFs link a centralized digital artefact repository (VF marketplace) with the tactical repository of VFs. 3) The tactical adversarial tactical cloud is susceptible to I-EDoS attacks, which may allow to fraudulent instantiate tactical VNFs on the edge. 4) The main adversarial repository is distributed into a pair of physical clusters. 5) One of the clusters contained the original VFs that supply the ISR capabilities to be requested, while the other contains Data Aggregation VNFs \cite{newTDoS99} that include a backdoor operable by the ally cyber commands that allow manipulate the information collected and processed at the edge. 
\item \textbf{Limitations and constraints}. Any physical actions (e.g. electronic warfare) against the adversarial tactical cloud, the digital supply infrastructure or their communication processes shall be conducted carefully enough so that the supply of the fraudulent VF is not affected. 
\item \textbf{Tactical end-state and objectives}. The end-state is achieved when the digital new ISR effectors deployed at the adversarial tactical cloud can be manipulated by the ally cyber command. The mission objective is to disrupt the forthcoming ISR action reported by human intelligence actors, thus preventing/deterring their subsequent kinetic infiltrations.
\end{itemize}

\begin{figure*}[t]
  \centering
  \includegraphics[width=0.6\linewidth]{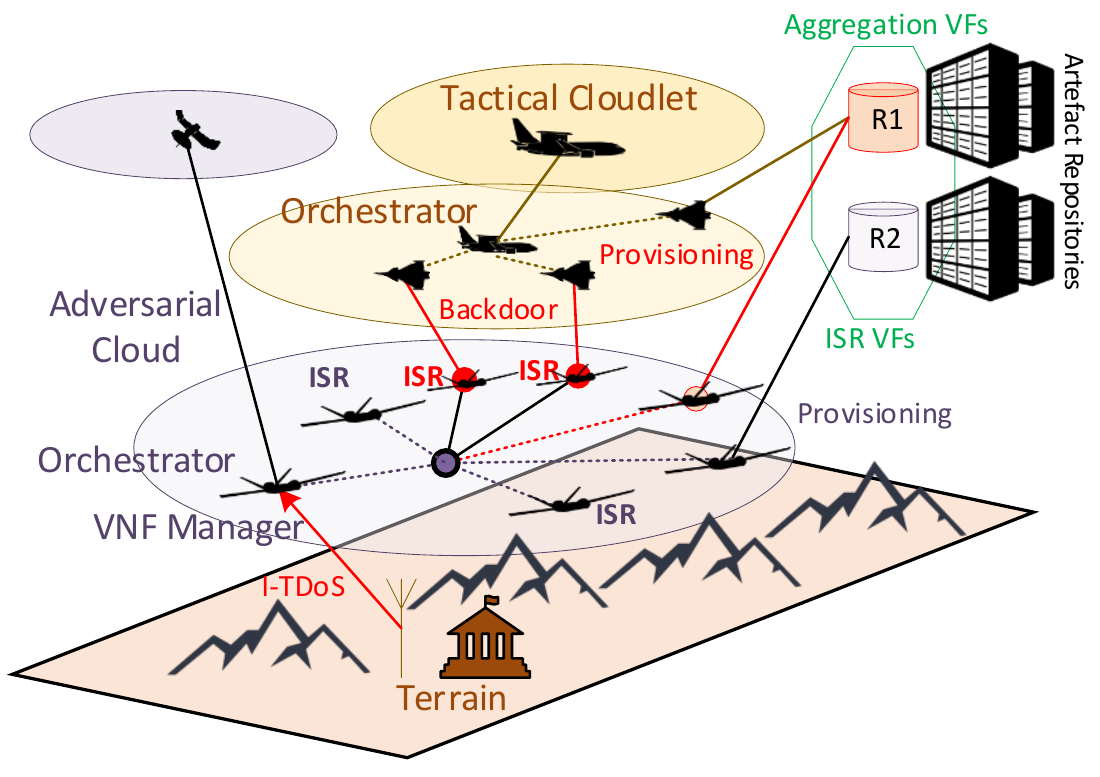}
  \caption{TDoS and the Digital Supply Chain}
  \label{conop5}
\end{figure*}

\noindent \textbf{Mission:} An operation between cyber commands unit operating fixed ICT infrastructure and the ally tactical cloud shall intercede in the adversarial ISR operations enabled by digital tactical capabilities, and from a hostile tactical cloud (see Fig. \ref{conop5}).

\noindent \textbf{Execution:} In order to deter the adversarial incursion, the Ally cyber command with enforce an offensive response action against the digital supply chain of the adversarial faction, for which the manipulation of the data provided by its new ISR capabilities is essential. In this context, the Ally cyber command will weaponize its fixed ICT capabilities towards firing an I-EDoS attack against the adversarial tactical cloud, which shall increase the computational cost of processing the multi-sourced ISR data. This shall lead to the adversarial VM Manager to request Data Aggregation VNFs \cite{newTDoS99} from a secondary VFs market, which stores jeopardized images for Data Aggregation. When the adversarial Orchestrator and VM Manager of the enemy tactical cloud instantiated them, the Ally cyber commands shall be able to exploit a backdoor on the corrupted Data Aggregators that allows forging the perceived information at the edge \cite{newTDoS100}. Consequently, the enemy will acquire feeds leading to a fraudulent operational picture manipulated towards persuading the infiltration.

\noindent \textbf{Tasks:} 1) Synchronization between the cyber command effectors available in range to the adversarial edge. 2) I-EDoS attack against the enemy VFN Orchestrator. 3) Wait for the enemy deployment of jeopardized data aggregation VNFs on their edge. 4) Exploitation of the backdoors on the jeopardized data aggregation VNFs. 5) Cyber sensed data poisoned aiming on infer a wrong operational picture. 6) Once the enemy infiltration is canceled, removal of evidences of the conducted cyber operation.

\noindent \textbf{Coordination instructions:} Information exchange on cyber and human intelligence, and the Ally cyber ISR actions for confirming both, the instantiation of jeopardized VNFs and the cancelation of the enemy infiltration, are distributed between the participants and effectors. The liaison, engagement rules, communication procedures and reporting systems are those comprised by the allied doctrines.

\noindent \textbf{Integrated support systems:} Cyber fires and cyber ISR are commanded from the ally cyber command HQ. Human intelligence is distributed thorough Federated Mission Networking (FMN). 

\begin{figure*}[t]
  \centering
  \includegraphics[width=0.6\linewidth]{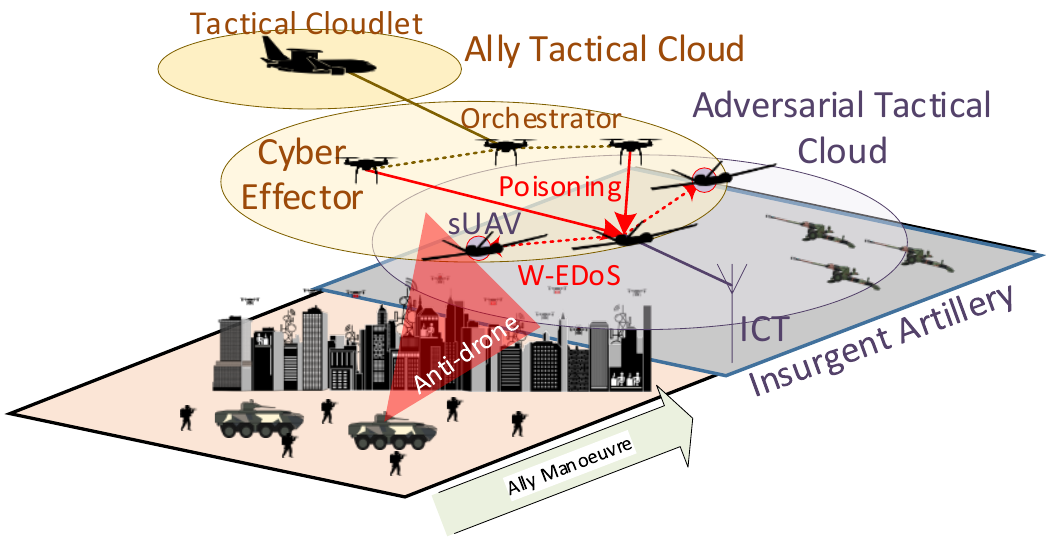}
  \caption{TDoS exposing the IR signature of adversarial drones}
  \label{conop6}
\end{figure*}

\subsection{CONOP\#5: Intensifying the IR signature of small hostile effectors}
\textbf{Background:} An ally faction will attempt to maneuver  around a metropolitan area dominated by hostile insurgents. Under the assumption of benefit from the surprise factor, it is expected that the targeted region will be mainly defenced by light infantry and few field artillery effectors. The insurgents are escorted by polyvalent small UAVs which digital payload adaptable and reconfigurable via NFV technologies; which allow customized recognition VFs onboarding as support to the deployed artillery effectors \cite{newTDoS101}. The VNF provisioning is enabled by a low intensity tactical cloud deployed at the insurgent edge on fixed ICT infrastructure.

\noindent \textbf{Current situation:} There is high asymmetry between the involved factions, where insurgents mostly focus on wear out and slow down the opponent with the expectation of taking advantage of side hybrid warfare actions. 
\begin{itemize}
\item \textbf{Own Forces}. Allies have superiority at all battle dimensions. Insurgents mostly rely on guerrilla tactics supported by proxy actor technology. 
\item \textbf{Co-operating and neutral actors}. Insurgents are technologically supported by proxy actors. Since the maneuvers will be conducted on a metropolitan area, there is a high likelihood of encountering civilian neutral actors on a complex urban terrain.
\item \textbf{Assumptions}. 1) If the insurgents discover the ally maneuvers, they will rely on their smalls UAVs for supporting their field artillery support fires. 2) UAVs operate on digital tactical services deployed as VNFs, and supplied by the adversarial tactical cloud. 3) Adversarial digital tactical services are susceptible to W-EDoS attacks. 4) The adversarial small UAVs reveal a very weak infrared (IR) signature, so they can barely been deactivated by the ally anti-drone effectors. 5) The higher the intensity of the VNFs workload, the higher IR signature visibility of the small UAV that carries them \cite{newTDoS102}. 6) The beast suitable way for the ally cyber command to reach the insurgent fixed ICT infrastructure is to deploy a tactical cloud on the operational metropolitan area. This cloud shall allow to instantiate and trigger the offensive digital tactical capabilities able to cause TDoS by W-EDoS on the adversarial services deployed at their edge. 
\item \textbf{Limitations and constraints}. The enforcement of jamming and denial of service related CoAs is not recommended by the analysts, since it may jeopardize the ally maneuver communications and impede the own cyber command operations. This also may derive on cyber EW collateral damage. 
\item \textbf{Tactical end-state and objectives}. The end-state is achieved when the kinetic maneuvers conclude. The mission objective is to disrupt the insurgent field artillery operations by exposing their support to targeting drones to the ally anti-drone systems.
\end{itemize}

\noindent \textbf{Mission:} Ally cyber commands shall support the urban kinetic maneuvers by facilitating the interception of the adversarial small UAVs. This will be conducted by triggering TDoS situations on the VFs they enable so their IR signature became more perceptible by the ally anti-drone effectors (see Fig. \ref{conop6}).

\noindent \textbf{Execution:} In order to support the ally maneuver, offensive digital tactical capabilities will be deployed on an urban tactical cloud. These capabilities shall allow cyber commands to poison the information exchanges between the fixed insurgent ICT infrastructure and the small UAV digital payload. The poisoning shall derived on W-EDoS \cite{tdos7}, so the CPU and memory usage of the general purpose processors of the drones intensify \cite{newTDoS101}, thus making their IR signature more detectable (temperature, emissivity, etc.) \cite{newTDoS102}. Then it is up to the ally anti-drone fires to intercept the small UAVs, the insurgent field artillery losing their support to targeting..

\noindent \textbf{Tasks:} 1) Synchronization between the cyber command and the kinetic deployment. 2) Identification of the insurgent fixed ICT infrastructure that enables their tactical cloud. 3) Deployment of offensive digital tactical capabilities able to reach the discovered infrastructure. 3) The cyber offensive effectors poison the telemetric and orchestration communications between the adversarial VF Manager and orchestrator and the VFs on the drones, leading to cause W-EDoS. 4) The ally IR sensors detect the small drones, since their workloads fraudulent intensified intensifying their signature. 5) The Ally anti-drone effectors neutralize the small UAVs via High Power Microwave (HPM) able to fire Electromagnetic Pulses (EMP) or guided ammunition \cite{newTDoS103}. 6) The cyber commands remove any evidence of the cyber offensive action.

\noindent \textbf{Coordination instructions:} Information exchange between cyber commands and the kinetic maneuvers for confirming the presence of hostile small UAVs and the effectiveness of the cyber actions on them. Anti-drone system shall confirm the effect of their fires at both cyber command and the kinetic maneuvers. The liaison, engagement rules, communication procedures and reporting systems are those comprised by the allied doctrines.

\noindent \textbf{Integrated support systems:} Cyber fires and maneuvers are commanded from the ally cyber command HQ. Cyber Threat Intelligence is distributed thorough Federated Mission Network (FMN). 

\section{Discussion}
The following discussed the key aspects, main opportunities and challenges reasoned from the CONOPs involving TDoS illustrated at Section 7.

\subsection{Rationale about CONOP \#1: jeopardizing the adversarial energy efficiency at ISR operations}
The scenario depicts a TDoS encounter aiming on disrupting the adversarial ISR actions enabled by hostile digital tactical capabilities deployed on an adversarial tactical cloud; in particular, VFs instantiable from a remote cyber asset inventory. The narrative of the CONOP introduces a hybrid situation, where a temporal peace agreement suggests that the opposite factions operations shall not be attributable (if possible not discoverable), in this way preventing an escalation of the conflict. The response of the ally faction against the enemy ISR actions resembles an Offensive Cyber defence Operation (OCO), where the attacker attempts to degrade the surveillance and recognition power of the adversarial faction. From a technical perspective, the allies may rely on both W-EDoS and I-EDoS attacks against the adversarial cloud, the first of them causing overload of CPU and Memory usage resulting in additional energy depletion; and the second causing overload in the VF Manager instantiation capabilities, thus resulting in energy depletion and the adversarial Cloudlets and data aggregation clusterheads. It is not expected direct socio-technical TDoS, although the presented mission will disrupt the adversarial capabilities for acquiring an accurate operational picture. 

From a tactical dimension, the narrative is mainly focused on justifying the opportunity of firing an M-TDoS situation on the adversarial tactical ISR capabilities, which is expected to lead the enemy decision-makers to enforce responsive CoAs to manage an unexpected energy depletion. The mission planner hypothesises that the enemy contingency plan will include to call back the effectors that carry the ISR sensor, so under a perfectly synchronization action, the ally artillery can replace its effectors. In general terms, the mission success depends on a wide set of premises (W-EDoS/I-EDoS able to jeopardize the enemy VFs, correct predictions on the adversary response, proper synchronization between ally cyber commands and artillery units, etc.), which makes it success difficult. 

As is usual in cyberspace operations, the mission will require a great preparation effort (simulation, vulnerability analysis, study of adversarial doctrine, etc.), but it will depend on fast execution times. Side conditions that have not been considered on the narrative despite entailing critical implementation challenges shall response questions like: how to assess the effectiveness of the involved cyber fires?, how ally forces will notice when the adversarial ISR effectors will be temporally out of the artillery units rage? Is there some potential cyber collateral damage? Or how to impede the attribution of the ally cyber command actions?. These questions can only be answered when digging into the particular operational conditions, which are out of the scope of the CONOP\#1 description. According to the mission narrative, it is assumed that joint commanders find feasible way of solving them.

\subsection{Rationale about CONOP\#2: TDoS against massive disinformation proxy tactics} 
The CONOP introduces mission that depicts a Civil-Military Co-operation (CIMIC) situations, where framed within a proxy warfare conflict, a faction supported by ally cyber commands shall take advantage of TDoS tactics in order to counter a massive disinformation campaign orchestrated by and adversarial tactical cloud. Given the proxy nature of the conflict, the ally cyber command shall operate undercovered, so it is essential that its implication on the TDoS fires must not be attributable. The CONOP describes a planned mission that define Internal Defensive Measures (IDM) able to deplete the adversarial capabilities for reaching the citizens on the targeted urban areas. The operational environment is a metropolitan area with fully functional 5G infrastructure and an operational civilian C-RAN layer, from which allies are able to effectively deploy defensive digital tactical capabilities. Based on this, defenders have a greater and more escapable CIS infrastructure, which shall be able to manage the hostile situation in properly weaponized. 

From the technical perspective, W-EDoS attacks are expected to be able to jeopardize the workload of the adversarial digital tactical capabilities while I-EDoS may reduce their potential of instantiate and smart distribute offensive VFs as the operational circumstances demand. However, they may potentially reveal the interference of the allies on a proxy faction, so the commander decided to rely on passive defensive actions: i.e. decoys, honeypots and sinkholes. From a tactical perspective, M-TDoS may contribute to reduce their workload. But the commander assumes that the most effective countermeasures will be based on deception triggering D-TDoS situations: it is assumed that the C-RAN infrastructure can better scale (in terms of connectivity, bandwidth, CPU usage, etc.) than the adversarial tactical infrastructure, so in an ideal case, to reach a one-to-one (hostile service – to – decoy situation) shall guarantee the mission success.

From the socio-technical perspective, the adversarial propaganda attempt to directly impact on the citizen perception of the conflict, but it will be very challenging to assess and evaluate the impact of the related enforced actions. The deception tactics operated by the defenders may misguide the attacker understanding of the social perceptions of the conflict. for example, they may lead them to think that the propaganda is being delivered, but that it is not causing any effect on the population, which can be discouraging and even negatively impacting on the morale of the attacker and its commitment to the cause. This may result in directly I-TDoS at both civilian and military dimensions while causing information chaos tentatively derived by misinformed decisions and failed communications (O-TDoS). In a higher level, this may vertically propagate as a rebound effect on the human capital life-cycle and HUMINT operations (E-TDoS). 

Overall, the CONOP\#2 is significantly easier to implement that the previous one, since its success does not heavily depends on strong hypothesis on the adversarial \textit{modus operandi} and technological conditions. It does not require such a great synchronization between ally cyber commands and related actions. However, it suffers the challenges inherent in CIMIC operations: most of the necessary infrastructure has mainly civilian use and industrial property, there is a large likelihood of causing cyber collateral damage, Rules Of Engagement (ROE) must consider strong civilian conditions, there is an increased risk of information leaks, some of the civil actors may not be sufficiently prepared to provide support in this type of scenario, etc..

\subsection{Rationale about CONOP \#3: TDoS fires supported by air combat operations}
The CONOP describes the mission plans for a join operation that makes converge an ally cyber offensive operation with combat air interception against and aerial air attack against an ally critical terrain. The presented mission outline a very challenging situation that demands a perfect synchronization and coordination of the involved effectors, otherwise potentially resulting in the loss of significant air assets (in the worse scenario, the loss of a critical terrain). In this context, it is expected that cyber commands take advantage of the exposure of attack surfaces on the adversarial digital tactical capabilities inherent to their resource reallocation for serving the drone swarm C2 and communications while dogfighting. The mission planner idea was to trigger a TDoS scenario that force the enemy to call back its effector due to battery loss earlier than expected, which requires that the ally air effectors maintain a dogfighting situation the time needed by the cyber commands to complete their action. 

Although the mission is very risky, given the air superiority of the attacker, the allies are left with few options for keeping their terrain. For example, the CONOP description assumes limitations concerning the application of EW Courses of Actions due to the full-spectrum criticality of the electromagnetic spectrum for operation-level communications. On the other hand, there are not indications about the availability of anti-drone systems, so it can be assumed that defenders lack these capabilities. The air interception operations cannot be sustained for long, so for a technical perspective, W-EDoS and I-EDoS fires shall convergently and strong impact on the exposed attack surfaces leading to M-TDoS and D-TDoS in all possible ways. From a socio-technical perspective, I-TDoS and O-TDoS impacts are expected, since it can be hypothesized that the unexpected depletion of the drone swam batteries may weaken the individual and operational trust on these enablers and the offensive operation that depends on them. This may affect decision-making, mislead the enemy lessons learned, and derive into E-TDoS cross-cutting impacts at all enterprise levels: order technical revisions/maintenance of the jeopardized assets, lack of trust on related industrial providers, questioning the preparedness of the operators and the awareness of the offensive decision-makers, etc.

\subsection{Rationale about CONOP \#4: Jeopardizing the Digital Supply Chain}
The proposed mission illustrates an alternative scenario to the CONOPs above that prompts the tactical advantages of TDoS when exploiting its synergies with offensive cyber actions against the cyber digital supply chain, the latter focalizing on the enemy capabilities of on demand provisioning their tactical services deployable at the edge as VFs. On these grounds, the scenario describes a symmetric conflict scenario where the allies attempt to plan and enforce fire TDoS fires able to reduce the adversarial ISR power, which is enabled by a distributed sensor grids orchestrated by a tactical cloud. The mission has a medium difficulty level, and its most challenging aspect depends on previous actions (preconditions); among them the need for preliminarily jeopardizing the VF images on the digital cornerstone that provisions the tactical operation, effectively weaponization of the poisoned VFs, cover operation of the cyber command that weaken the detection and attribution of the remote ally cyber operations, trust in the attacker not discovering the data/processing forgeries and exploiting them in terms of counterintelligence, or certainly on the VF Manager/orchestrator instantiation Data Aggregation services against ISR sustainability issues. The CONOP indicates limitations concerning the use of EW tactics against the adversarial tactical cloud, so all the cyber actions will consider from data link layer (OSI model Layer 2) upwards. Collateral damage may be mainly due to the exposure of sensitive information, and the instantiation of the jeopardized VFs in contexts beyond the scope of the described operation (which may include civilian operations, especially if the cyber asset inventory is linked to dual-use repositories).

From the technical perspective, the mission describes a principal TDoS action to be enforced by the ally cyber command: to fire undercovered W-EDoS and I-EDoS against the adversarial ISR VFs until the VF Manager and Orchestrator decide to call supportive Data aggregation services. The expected effect of W-EDoS is to increase (not denying) the consumption of computation resources per VNF, while I-EDoS shall fraudulently increase the horizontal escalation of the services making grow the ISR expenditures via upkeep of “lazy” services. From a technical perspective, the first will be linked to M-TDoS on the adversarial digital tactical capabilities while the second will vertically propagate a D-TDoS situation. 

Since the counterintelligence operation attempts to disrupt the capability of the enemy of perceiving the operational picture, the presence of I-TDoS implications in inherent, which among others may include jeopardizing the individual capabilities for acquiring situational awareness, reducing the personnel trust on the supportive tactical cloud and its sensor grid, or triggering cognitive/social disruptions by forgery and injection of sensitive information and propaganda. The related O-TDoS impacts are linked to the quality of the operational information exchanges and decisions, and the E-TDoS consequences will derive on the claim of additional enterprise-level needs (additional preparedness, replacement of effectors, difficulties in further recruiting campaigns, etc.) and diverse hybrid threats (politics economics, social, etc.)

\subsection{Rationale about CONOP \#5:  Intensifying the IR signature of small hostile effectors}
The presented scenario depicts an asymmetric conflict where allies have superiority at all battle dimensions, which is mostly negated by an urban warfare confrontation against the adversarial insurgent faction. The mission plan reveals that insurgents are supported by field artillery, which is able to weaponize recognition small UAVs for guiding its fire life-cycle. The ally commander relies on a heavy assumption about the side-impact of deploying/maintaining digital tactical services at the edge: that a significant increase or oscillation on the VFs CPU/Energy efficiency will make more visible their IR signature. This hypothesis is supported by state-of-the-art publications and simulation, so it has sense to infer that TDoS fires may offensively trigger these variations, resulting in a tactical advantage for the ally anti-drone effectors. The mission describes a related cyber offensive operation as support to a kinetic deployment  via reducing the adversarial field artillery power.

The CONOP does not provides evidence about how UAVs may support artillery (EO/IR target identification, damage assessment, etc.) nor if side-channels attacks derive from the inferred electromagnetic emanations may reveal additional information beyond the sole IR signature, the later tentatively enabling interesting related courses of action. Although direct EW actions against them are not supported because of the assumption of potentially damaging the maneuver tactical communications, collateral damage may impact on civilian infrastructure (cyber), and "false positives" at IR signature recognition may derive in fires against neutral targets, resulting in physical, logical or even personal damage and their social implications. The CONOP does not state that the cyber offensive shall be enforced undercover, but this is recommended in order to thwart the lessons learned by the enemy, which for example may have repercussions on the surprise factor at future related actions.

The mission entails medium difficulty mostly due to large preparatory dependencies (simulation and modelling of IR signatures of VFs, enumeration of attack surfaces and their exploitation for TDoS purpose, etc.), the difficulty inherent on operating an ally tactical cloud on a metropolitan region, and the reachability of the cyber targets.  From a technical perspective, W-EDoS attacks will hardly focus on disrupting the VFs workload sustainability while I-EDoS attacks may only target communication cluster heads, since the instantiation of ``lazy” nodes may create confusion and difficult the IR signature recognition. In analogy, M-TDoS shall be vertically propagated from the technical plane due to W-EDoS impacts, while D-TDoS will be linked to D-TDOs. The mission does not describe direct socio-technical goals, but significant side effects can be expected. For example, I-TDoS may be featured by individual lack of trust on the supportive small UAVs, confusion by not knowing why their assets are being detected or demotivation as the enemy infantry can’t be properly supported by their field artillery. The organizational side impacts (O-TDoS) will be mainly linked to inaccurate decision making and a wrong understanding of the opposite faction capabilities. At enterprise-level, most E-TDoS consequences will be related to wrongly involving preparedness and capacitation actions.

\section{Conclusions}
This paper presented the evolution of the conventional denial of sustainability attacks towards the emerging Tactical Clouds, highlighting their typology, potential attack surfaces and impacts when released on the military edge. In this context, the concept Tactical Denial of Sustainability (TDoS) have been developed, including two raw technical variations: Maintenance-based Tactical Denial of Sustainability (U-TDoS) and Deployment-based Tactical Denial of Sustainability (D-TDoS); and their horizontal/vertical socio-technical propagations as Enterprise (E-TDoS), Organizational (O-TDoS) and Individual (I-TDoS) levels. The paper also revealed the close synergy between TDoS and energy efficiency situations, as well as how the later may jeopardize the tactical availability of tactical clouds. 

Their potential causality and impact on tactical military operations have been highlighted by five illustrative Concepts of Operation (CONOPS), covering their weaponization on hybrid, proxy and symmetrical operations. Given the important gap in the bibliography on related topics, the high level of secrecy surrounding the Tactical Cloud paradigm, and the novelty of the technological ecosystem in which it relies on; the scope of this publication has been to introduce the aforementioned topics to both military and general public with the clear purpose of incenting further research and further exploring the potential of the Tactical Cloud concept in the modern conflict scenarios. Because of this, many aspects of particular interest to the authors have not been treated as thoroughly as they would have liked, including: to describe in more detail the related ICT enablers and their inter dependencies, to explain related vertical/horizontal TDoS propagation assessment capabilities, or to in-depth review the potential tactical and CIS Courses of Action (CoAs) against adversarial actions trying to jeopardize the sustainability of the digital tactical assets. They will be developed as future research, and relaying on particular military capacitation.

\section*{Disclaimer}
The contents reported in the paper reflect the opinions of the authors and do not necessarily reflect the opinions of the respective agencies, institutions or companies.

%

\bibliographystyle{ACM-Reference-Format}
\bibliography{sample-base}

\end{document}